\documentclass{WileyChemistry-template}

\usepackage{lipsum}
\usepackage{comment}
\usepackage{xcolor}

\title{\raggedright Photophoretic Trapping: Fundamentals, Advances and Future Directions}

\author{
\begin{minipage}{\textwidth}
	Anita Pahi,\textsuperscript{[a]} Kirty Ranjan Sahoo,\textsuperscript{[a]} Souvik Sil \textsuperscript{[b]}  and Ayan Banerjee\textsuperscript{*[a]} 
\end{minipage}
}

\newcommand{\affiliation}{
\begin{itemize}

\item[{[a]}] Indian Institute of Science Education and Research Kolkata, 741246, India\\
E-mail: ayan@iiserkol.ac.in\\

\item[{[b]}] Okinawa Institute of Science and Technology Graduate University (OIST), 1919-1 Tancha, Onna-son, Okinawa, Japan 904-0495\\

\end{itemize}
}

\renewcommand{\abstract}{Photophoretic forces, several orders of magnitude stronger than radiation pressure, enable particle trapping at remarkably low optical intensities and have opened pathways to applications in aerosol science, free-space 3D volumetric displays, and even deployment of lightweight payloads in space. In this review, we provide a comprehensive explanation of the underlying physics of photophoretic forces and how they facilitate stable three-dimensional manipulation of absorbing particles. We examine the experimental configurations that enable robust trapping, and we detail the physical parameters that govern the magnitude and behavior of photophoretic forces in these geometries. The rich dynamical phenomena exhibited by photophoretically trapped particles are discussed alongside current and emerging applications and possible future research directions. This review thus attempts to systematically unify the theoretical, experimental, and application-oriented aspects of photophoretic trapping, with the aim of advancing and strengthening research in this rapidly developing field.
}

\newcommand{\keywords}{
	Photophoretic force \textbullet\ 
	$\Delta T$ and $\Delta \alpha$ photophoretic force\textbullet\ 
	Fluctuation dynamics \textbullet\ 
	Brownian motion \textbullet\ 
	Aerosol \textbullet\
    photophoretic flight
}

\begin{document}

\twocolumn[ \vspace{-1.5cm}\maketitle\vspace{-1cm}
	\textit{\dedication}\vspace{0.4cm}]
\small{\begin{shaded}
		\noindent\abstract
	\end{shaded}
}

\begin{figure} [!b]
\begin{minipage}[t]{\columnwidth}{\rule{\columnwidth}{1pt}\footnotesize{\textsf{\affiliation}}}\end{minipage}
\end{figure}




\section*{Introduction}

The ability to confine a single aerosol particle in space provides a powerful platform for investigating its intrinsic physical and chemical properties over extended periods, free from external disturbances\cite{dettlaff2025trapping,zhang2023resonance}. Optical trapping has thus emerged as a vital technique for high-precision single-particle studies. Moreover, such techniques have found important applications in vacuum environments, where they are used to explore quantum phenomena at the microscopic scale\cite{chan2011laser,ma2020observation,iakovleva2023zeptometer,kamba2023revealing}. Optical trapping in liquids has been facilitated quite robustly by the advent of single mode, reasonably high power lasers which have been used to develop `Optical Tweezers'\cite{ashkin1986observation}, for which Arthur Ashkin was awarded the Nobel Prize in 2018. The issue becomes more complicated in air, however, since the viscosity of air is much lower than liquids, leading to large diffusion lengths of Brownian particles which thus require steep potential wells to confine them. Thus, in air, two primary mechanisms are employed to achieve optical trapping: radiation pressure force and photophoretic force. Radiation pressure, which is well understood, is effective for levitating transparent particles. On the other hand, photophoretic force -- that is of great efficacy in trapping absorbing and irregularly shaped particles -- is more complex and less thoroughly characterized. Nevertheless, ongoing experimental advancements are gradually deepening our understanding of photophoretic trapping, though its full potential is yet to be realized. With this brief background, we now turn to the historical development of optical trapping in air.
\label{introduction}

\subsection*{A Brief Look into the Historical Background}
The first observation of particle stabilization using light was reported in 1966 inside a laser cavity, where the particles involved were simply dust \cite{rawson1966propulsion}. Calculations here showed that the dominant force responsible for stabilization was the photophoretic force, rather than radiation pressure force --  whose magnitude was found to be significantly smaller. 

Subsequently, in 1971, optical levitation in air was first demonstrated by Arthur Ashkin using the principle of radiation pressure alone~\cite{Ashkin1971}. In his experiments, transparent particles were used specifically to avoid the effects of photophoretic forces. The combined effect of scattering and gradient forces directed along and transverse to the trapping beam axis enabled the levitation of transparent particles in air. 

Ashkin also showed that at reduced pressures, the levitated particles tended to fall, which he attributed to increasing thermal forces (i.e., photophoretic effects) that become more significant as the pressure dropped below atmospheric levels. This highlighted a critical limitation: in the configuration Ashkin employed, radiation pressure alone was insufficient for trapping absorbing particles, because for such particles, the photophoretic force is approximately $10^{5}$ times stronger than the radiation pressure force\cite{lewittes1982radiometric}. 

In 1982, Lewittes demonstrated stable levitation of absorbing particles in air using photophoretic forces, marking a significant development in optical trapping methods for absorbing particles\cite{lewittes1982radiometric}. Here, he achieved the radiometric levitation of a \(20\,\mu\text{m}\) diameter absorbing particle at an intensity of approximately \(1\,\text{W/cm}^2\), in contrast to the \(20\,\mu\text{m}\) diameter transparent sphere used by Ashkin, which required an intensity of approximately \(10^5\,\text{W/cm}^2\) to achieve levitation via radiation pressure. Since thermal forces are significantly stronger than radiation pressure forces, the intensity required to trap absorbing particles is considerably lower.  Lewittes employed a $TEM_{01}$ beam profile for this purpose. Subsequently, Pluchino~\cite{pluchino1983radiometric} demonstrated stable radiometric levitation using a $TEM_{00}$ beam profile, showing that photophoretic trapping can also be realized with a Gaussian beam profile. We now turn to a basic understanding of the photophoretic forces.

\subsection*{Theory of Photophoretic forces}

Photophoretic forces arise from the interaction between light and an absorptive particle that becomes hotter than its surrounding gas environment due to this interaction. For this force to manifest, the particle must exhibit surface inhomogeneity either in temperature distribution or in the gas accommodation properties. When gas molecules collide with the particle surface and get reflected, they carry away more momentum from the hotter regions than from the cooler ones. This asymmetric momentum exchange generates a net recoil force on the particle, known as the \textit{photophoretic force} \cite{horvath2014photophoresis,jovanovic2009photophoresis}. A classic demonstration of this principle is the well-known Crookes radiometer invented by Sir William Crookes, see figure \ref{delta_T_delta_Alpha}(a). This device consists of four vanes mounted on a spindle inside a partial vacuum bulb, with each vane painted black on one side and white (or silvered) on the other. When exposed to light, the vanes rotate such that the the white surfaces advance in front of the black, i.e., the rotation is clockwise when viewed from above. This occurs because the black surfaces absorb more light, become hotter, so that the net photophoretic force is directed from the black to the white side\cite{horvath2014photophoresis}. Notably, if the device operated solely on radiation pressure, the rotation direction would be opposite, as radiation pressure exerts more force on the reflective (white) surfaces. When the surface temperature of a particle has a temperature \( T_s \), which is higher than the surrounding gas temperature, \( T_i \), gas molecules that collide with the hot surface can absorb thermal energy and are subsequently reflected with velocities corresponding to a higher effective temperature \( T_r \),  greater than \( T_i \).  
The efficiency of this thermal energy transfer is quantified by the \textit{thermal accommodation coefficient} \( \alpha \), defined as: 
\begin{equation}
   \alpha = \frac{T_r - T_i}{T_s - T_i}
\end{equation}

Here, \( \alpha = 0 \) corresponds to no energy exchange (perfectly elastic reflection), and \( \alpha = 1 \) indicates full thermal accommodation, where gas molecules completely equilibrate with the surface temperature. The value of \( \alpha \) depends on various factors, including the material composition and surface morphology of the particle, as well as the type of gas in the environment. Importantly, the photophoretic force is defined on a particle when the accommodation coefficient($\alpha$) on the particle is non zero. The force can be ideally divided into two types- (i) The photophoretic $\Delta T$ force and the photophoretic $\Delta \alpha$ force.

\subsubsection*{Photophoretic $\Delta T$ force}

\begin{figure*}[t]
\centering
\includegraphics[width=\linewidth]{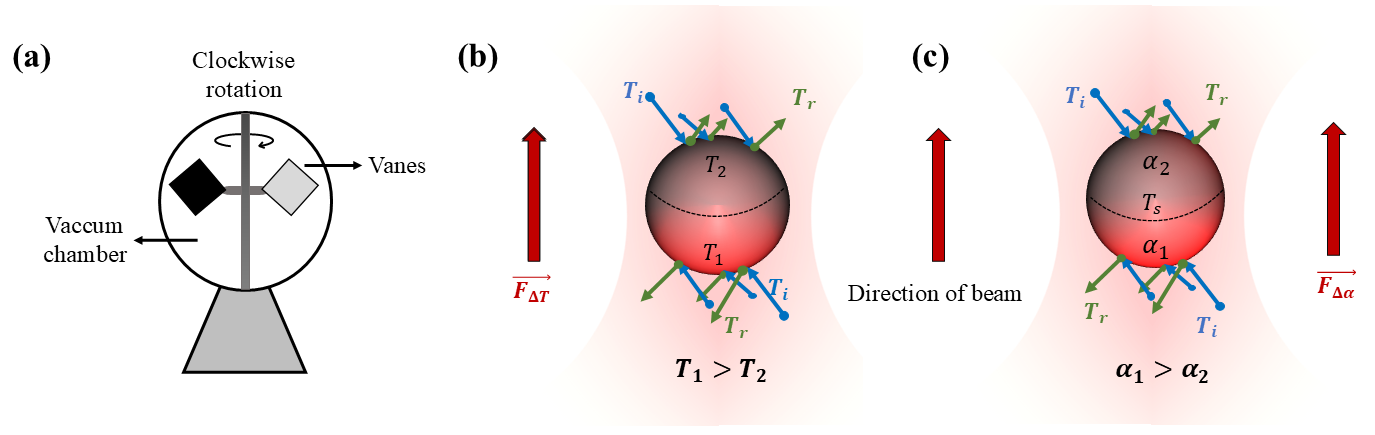}
\caption{ (a) Schematic of a Crookes radiometer (b) and (c) show schematic representations of the photophoretic forces due to a temperature gradient (\(\Delta T\)) and an accommodation coefficient gradient (\(\Delta \alpha\)), respectively. The left arrows indicate the direction of the \(\Delta T\) photophoretic force in (a) and \(\Delta \alpha\) in (b). The central arrow represents the direction of the light beam incident on the particle.}
\label{delta_T_delta_Alpha}
\end{figure*}

The $\Delta T$ photophoretic force(denoted as $F_{\Delta T}$) can be explicitly understood by assuming a constant accommodation coefficient across the particle surface. The interaction of an absorbing particle with light results in one side of the particle becoming warmer than the other. This leads to a temperature difference across the particle surface. As a result, the gas molecules hitting the particle from the warmer side leave with a temperature which corresponds to a higher velocity compared to the ones hitting the colder side of the particle. This leads to an asymmetric momentum transfer to the particle and a resultant force on the particle which is directed from the higher to the lower temperature. For a strongly absorbing particle, the illuminated side of the particle becomes hotter than the opposite side -- as a result of which, the force acts from the hotter (illuminated) side to the cooler side along the direction of the incident light and is referred to as the \textit{positive photophoretic force} as shown in figure \ref{delta_T_delta_Alpha}(b). However, in the case of weakly absorbing particles, that are partially transparent as a result,  the rear (non-illuminated) side becomes warmer, with the incident light being directed towards that side. This is also called  the lensing effect\cite{horvath2014photophoresis}. In such cases, the resulting force acts opposite to the direction of illumination and is termed the \textit{negative photophoretic force}.  
Since the temperature gradient, and hence the direction of the force, is determined by the direction of the incident beam, this type of force is referred to as a \textit{beam-fixed} force.

\subsubsection*{Photophoretic $\Delta \alpha$ force}\label{alpha_force}

Even in the absence of a temperature gradient across the particle surface, i.e. when the particle is uniformly heated and has a constant temperature \( T_s \) across its surface (refer to Fig.~\ref{delta_T_delta_Alpha}(c)), a net force can still act on the particle if there exists a variation in the accommodation coefficient across its surface. Such a condition can reasonably arise in the case of irregular or asymmetric particles, where the asymmetry may originate from differences in shape, surface texture, or internal density distribution. In these scenarios, gas molecules interacting with regions of the surface that possess a higher accommodation coefficient    acquire more thermal energy and are reflected with greater velocities than those interacting with regions of lower accommodation. This asymmetry in momentum transfer gives rise to a net photophoretic force on the particle, known as the photophoretic \( \Delta \alpha \) force(denoted as $F_{\Delta \alpha}$ ). For simplicity, consider a spherical particle where the lower hemisphere has a higher accommodation coefficient (\( \alpha_1 \)) and the upper hemisphere has a lower one (\( \alpha_2 \)). The resulting force acts from the region with higher accommodation (\( \alpha_1 \)) toward the region with lower accommodation (\( \alpha_2 \)), as shown in (Figure \ref{delta_T_delta_Alpha}(c)). Since this force depends on the particle’s material and structural properties, its direction varies with the particle’s orientation. Therefore, it is referred to as the \textit{body-fixed} force. 

\subsubsection*{Mathematical expression for $F_{\Delta T}$ and $F_{\Delta \alpha}$}
The calculation of the photophoretic force depends on the regime in which the particle resides, determined by the Knudsen number (\(K_n\)). The Knudsen number is defined as the ratio of the mean free path of the gas molecules (\(\lambda\)) to the particle radius (\(a\)), i.e., \(K_n = \tfrac{\lambda}{a}\). Based on the value of \(K_n\), three distinct regimes are identified as:\\ 
(i) the free-molecular regime (\(K_n \gg 1\)), also called the low pressure regime,  (ii) the continuum regime (\(K_n \ll 1\)), also called the high pressure regime and
(iii) the intermediate or the transition regime (\(K_n \approx 1\)).

\textit{\textbf{$F_{\Delta T}$ force :}} The calculation of $F_{\Delta T}$ force is done assuming that the accommodation coefficient $\alpha$ is constant and the temperature at the surface of the particle is rotationally symmetric since this is a space-fixed force. The expression for photophoretic force for the free molecular regime($F^{fm}$) and for the continuum regime($F^{co}$) is given as follows\cite{rohatschek1995semi, horvath2014photophoresis}:\\
\begin{equation}
    F^{fm} = 2D \, \frac{p}{p^*} \, a \, \frac{\alpha}{2} \, A_1
\end{equation}

\begin{equation}
    F^{co} = 2D \, \frac{p^*}{p} \, a \, A_1
\end{equation}

Here, $D$ denotes a constant determined entirely by the state of the gas,

\begin{equation}
    D = \frac{\pi}{2} \sqrt{\frac{\pi}{3}} \, \kappa \, \frac{\bar{c}\,\eta}{T_\infty},
\end{equation}

whereas $p^*$, which has the meaning of a characteristic pressure, also depends on the particle radius:

\begin{equation}
    p^* = \frac{1}{2} \sqrt{3\pi \kappa} \, \frac{\bar{c}\,\eta}{a}
    = \frac{3}{\pi} D \, \frac{T_\infty}{a}.
\end{equation}
Here, a = particle radius, $\kappa$ = thermal creep coefficient (sometimes called the slip-flow coefficient), $T_{\infty}$ denotes the temperature of the gas far from the sphere, $A_1 = \frac{1}{2} \Delta T_s$ with $\Delta T_s$ being the difference in temperature in particle surface. $\bar{c} \, \left( = \sqrt{\tfrac{8RT}{\pi M}} \right)$ is the mean velocity of the gas molecules, 
where $R$ is the universal gas constant, $T$ is the temperature, and $M$ is the molecular weight of the gas molecules. One significant difference between the force expressions in the continuum and free-molecular regimes is that the $\Delta T$ force is directly proportional to pressure in the former and inversely proportional to pressure in the latter.\\

The photophoretic force in the intermediate or the transition regime($F$) is calculated using the interpolation scheme proposed by Hettner\cite{hettner1926theorie}\\
\begin{equation}
    \frac{1}{F} = \frac{1}{F^{fm}} + \frac{1}{F^{co}}
\end{equation}

\textit{\textbf{$F_{\Delta \alpha}$ force :}}

The calculation of the photophoretic \( \Delta \alpha \) force is based on the difference in thermal accommodation coefficients across the particle surface, which can arise from variations in surface roughness or composition. Mathematically, the $\Delta \alpha$ force is expressed in terms of a function $\phi$, which depends only on the gas properties and the first-order Legendre coefficient $B_1$. The coefficient $B_1$ is the first order Legendre coefficient of $T_a$, the temperature of the gas layer adjacent to the particle surface and quantifies the degree of asymmetry in the accommodation-coefficient distribution around the sphere.  Two equivalent formulations of the force exist, depending on how the temperature difference \((T_s - T_i)\) is expressed:  

\begin{equation}
F_{\Delta \alpha} = \phi \, B_1 = \frac{3}{4} D \, a \, \frac{p}{p^\ast + \tfrac{p^\ast}{p}} (T_s - T_i)\,\Delta \alpha,
\label{eq:DeltaAlpha1}
\end{equation}

\begin{equation}
F_{\Delta \alpha} = \frac{1}{12}\, \frac{1}{\bar{c}}\, \frac{1}{1 + \left(\tfrac{p}{p^\ast}\right)^2}\, \frac{AI \, \Delta \alpha}{\alpha},
\label{eq:DeltaAlpha2}
\end{equation}

\begin{equation}
F = \frac{1}{2 \cdot \bar{c}} \cdot 
    \frac{\gamma - 1}{\gamma + 1} \cdot 
    \frac{1}{1 + \left( \frac{p}{p^*} \right)^2} \cdot 
    \frac{\Delta \alpha}{\alpha} \cdot H
\label{eq:DeltaAlpha3}
\end{equation}

where \(A\) is the absorption cross-section, \(I\) is the incident intensity, \(\alpha\) is the mean accommodation coefficient, \(p^\ast\) is the characteristic pressure, $\gamma$ = specific heat ratio of the surrounding gas, and $\Delta \alpha $ is the difference in accommodation coefficient across the particle surface and \(\bar{c}\) is the mean molecular velocity.  

Thus, depending on whether the formulation is based on the direct surface temperature difference \((T_s - T_i)\) or the heat flux exchange \(H\), one may use Eqs.~\eqref{eq:DeltaAlpha1}--\eqref{eq:DeltaAlpha3} to estimate the photophoretic \(\Delta \alpha\) force\cite{rohatschek1995semi}. 

Having established an understanding of the photophoretic \( \Delta T \) and \( \Delta \alpha \) forces, we will discuss in detail in following sections, how these forces enable three-dimensional (3D) optical trapping of absorbing particles in air.

\subsection*{Particles employed in Photophoretic trapping}
For the photophoretic force to manifest, the particle must absorb light at the trapping laser wavelength, either strongly or weakly. Additionally, particle asymmetry gives rise to the $\Delta \alpha$ force, which plays a crucial role in confinement, as discussed in the following sections. A variety of strongly absorbing particles have been successfully trapped, including single- and multi-walled carbon nanotubes (SWCNTs, MWCNTs), Bermuda grass smut spores, Johnson grass smut spores, natural graphite nanopowders, HB-2 pencil lead powder, CuO microparticles, B$_4$C particles, carbon printer toner, and carbon black \cite{gong2016optical, pahi2024study, pahi2025light, bera2016simultaneous, wang2015optical, chen2018temporal, lin2017measurement, sil2020study, lin2014optical} . In contrast, weakly absorbing particles such as English oak pollen and paper mulberry pollen have also been reported to remain stably trapped \cite{gong2016optical}. The size of trapped particles spans from a few micrometres up to nearly 100 $\mu m$. Most of these particles exhibit asymmetry in shape, which supports the action of the $\Delta \alpha$ force. Nonetheless, there are also reports of spherical absorbing particles, including iron microparticles and ink droplets, graphite particles, being trapped by photophoretic forces \cite{liu2014photophoretic, he2019investigation}, though these appear to be far more difficult to trap compared to asymmetric particles. However, it also appears that a complete quantitative understanding of how surface morphology asymmetry affects photophoretic force-based trapping has yet to be developed. 

Let us now consider how photophoretic forces enable three-dimensional (3D) optical confinement of absorbing particles in air across various experimental configurations.

\subsection*{Principle of Trapping Using Photophoretic Force}
Various optical configurations have been employed for trapping absorbing particles in air -- namely the \textit{vertical}, \textit{horizontal}, and \textit{counter-propagating beam} geometries. Trapping in these geometries results from the combined effects of photophoretic forces associated with surface temperature gradients ($F_{\Delta T}$), non-uniform thermal accommodation ($F_{\Delta \alpha}$) of the particle and gravity. 
Here, we discuss the intuitive physical principles that have been proposed in the literature to explain the trapping mechanisms in these configurations.

\subsubsection*{Horizontal Configuration}
In the horizontal configuration, a single focused Gaussian beam has been employed to achieve photophoretic trapping of particles in air \cite{gong2016optical, lin2014optical, zhang2012observation, he2019investigation}. In this geometry, the Gaussian beam propagates along the horizontal $z$-axis (Figure \ref{trap_mech}(a)). The photophoretic force arising due to the temperature gradient across the surface of the particle ($F_{\Delta T}$), i.e. the space-fixed force -- dependent solely on the direction of incident light -- acts along the beam propagation direction ($z$-axis). In contrast, the gravitational force ($F_G=mg$) acts vertically downward along the $y$-direction, which is transverse to the beam propagation direction. The photophoretic force arising from the gradient in thermal accommodation coefficient (\( F_{\Delta \alpha} \)) plays a crucial role in stabilising the particle in this configuration. Depending on the net force balance, \( F_{\Delta \alpha} \) tends to orient itself opposite to the combined direction of \( F_{\Delta T} \), \( F_R \), and \( F_G \). When \( F_{\Delta \alpha} > F_{\Delta T} + F_R + F_G \), the particle moves toward the beam focus, thereby increasing \( F_{\Delta T} \) and \( F_R \), until stability is achieved at:
\begin{equation}
    F_{\Delta \alpha} = F_{\Delta T} + F_R + F_G.
    \label{horizontal-trap}
\end{equation}
Conversely, if \( F_{\Delta \alpha} < F_{\Delta T} + F_R + F_G \), the particle moves away from the focal point, reducing the optical forces until the same balance is restored. This dynamic adjustment mechanism has been intuitively described in the work of Ze Zhang \textit{et al.} \cite{zhang2012observation} Building upon this, \textit{J. Lin et al.} \cite{lin2014optical}, in their study on the direct measurement of periodic circular motion of individual micron-sized absorbing particles in air trapped by a single focused Gaussian beam ($\mathrm{TEM}_{00} $), provided a more nuanced trapping mechanism. They decomposed the $ F_{\Delta \alpha} $ into its \textit{longitudinal} ($z$-axis) and \textit{transverse} ($x, y$) components to elucidate their observed circular motion. The longitudinal component $ F_{\Delta \alpha}^{(z)} $ balances the axial forces $ F_{\Delta T} + F_R $, ensuring axial stability. Meanwhile, the transverse components $F_{\Delta \alpha}^{(x,y)} $ serve a dual function: they counterbalance the gravitational force in the vertical direction and simultaneously provide the necessary \textit{centripetal force} and \textit{torque} required for the observed sustained \textit{circular motion}. The overall mechanism has been illustrated in the Figure (\ref{trap_mech}(b)). A comparative study of vertical and horizontal photophoretic trapping using both Gaussian and hollow beams examined the behaviour of regular and irregular particles~\cite{he2019investigation}. Their results showed that regular (symmetric) particles cannot be stably trapped in the horizontal configuration, as the photophoretic force due to the gradient in thermal accommodation coefficient ($F_{\Delta \alpha}$) vanishes, leaving only $F_{\Delta T}$ and gravity, which act orthogonally and therefore cannot establish a stable force balance. In contrast, for asymmetric particles, one component of $F_{\Delta \alpha}$ counteracts gravity while the other balances $F_{\Delta T}$, enabling stable two-dimensional confinement in the horizontal configuration.



\subsubsection*{Vertical Configuration}
Having understood the trapping mechanism in horizontal geometry, we now turn our attention to the vertical configuration, which is another widely used scheme in the photophoretic air trapping community \cite{sil2020study, bera2016simultaneous, pahi2025light, pahi2024study}. While the fundamental forces involved in trapping remain the same, the beam propagation direction opposite to the gravity introduces important differences in how stable trapping is achieved.
\par 
In the vertical configuration, a single focused Gaussian beam propagates along the vertical ($+z$)-axis (Figure \ref{trap_mech}(c)). As previously discussed, both the photophoretic force arising from a temperature gradient across the particle surface ($F_{\Delta T}$) -- being a space-fixed force -- and the radiation pressure force ($F_R$), act along the beam propagation direction ($+z$-axis). In contrast, the gravitational force ($F_G = mg$) acts downward, opposing these two forces. Similar to the horizontal configuration, the photophoretic $\Delta \alpha$ force plays the critical role in achieving both axial and radial confinement. This force, being body-fixed, does not have a fixed global direction but rather orients at an angle relative to the vertical axis depending on the particle’s asymmetry, orientation, surface roughness and other morphological properties. On decomposing this force into its axial and radial components, the axial trapping is achieved through the balance:
\begin{equation}
    F_{\Delta \alpha}^{(z)} + F_{\Delta T} + F_R = F_G
    \label{vertical-trap}
\end{equation}
In the radial direction, confinement is governed solely by the transverse components of the body-fixed force, $F_{\Delta \alpha}^{(x,y)}$. Analogous to gradient forces in conventional optical traps, the transverse components act as a purely restoring force\cite{bera2016simultaneous}. Importantly, there are no opposing forces in the radial direction, hence the restoring action of $F_{\Delta \alpha}^{(x,y)}$ becomes essential for radial trapping. The interaction of $F_{\Delta \alpha}^{(x,y)}$ with the gravity creates a rotational motion. Concurrently, the longitudinal component $F_{\Delta \alpha}^{(z)}$ gives another torque, resulting in a complex three-dimensional rotational motion of the trapped particle \cite{sil2020study}. Notably, this complex rotational motion leads to radial confinement in such traps. This overall mechanism has been illustrated in Figure (\ref{trap_mech}(d)). Interestingly, there are particles that exhibit no rotation, which raises interesting questions about the underlying mechanisms governing their stability in the trap. This behavior can be attributed to the particle morphology, which likely aligns the photophoretic force primarily along the vertical direction, while the gradient force provides the necessary radial restoring effect. This interpretation is supported by observations showing that non-rotating particles experience an effective viscous drag in the transverse direction significantly higher than that of ambient air \cite{pahi2025light}, which -- very crucially -- may account for the stable radial confinement observed for particles trapped by a single, loosely focused Gaussian beam in the vertical configuration.
\subsubsection*{Counter propagating beams Configuration}
\begin{figure*}[t]
    \centering
    \includegraphics[width=0.9\textwidth]{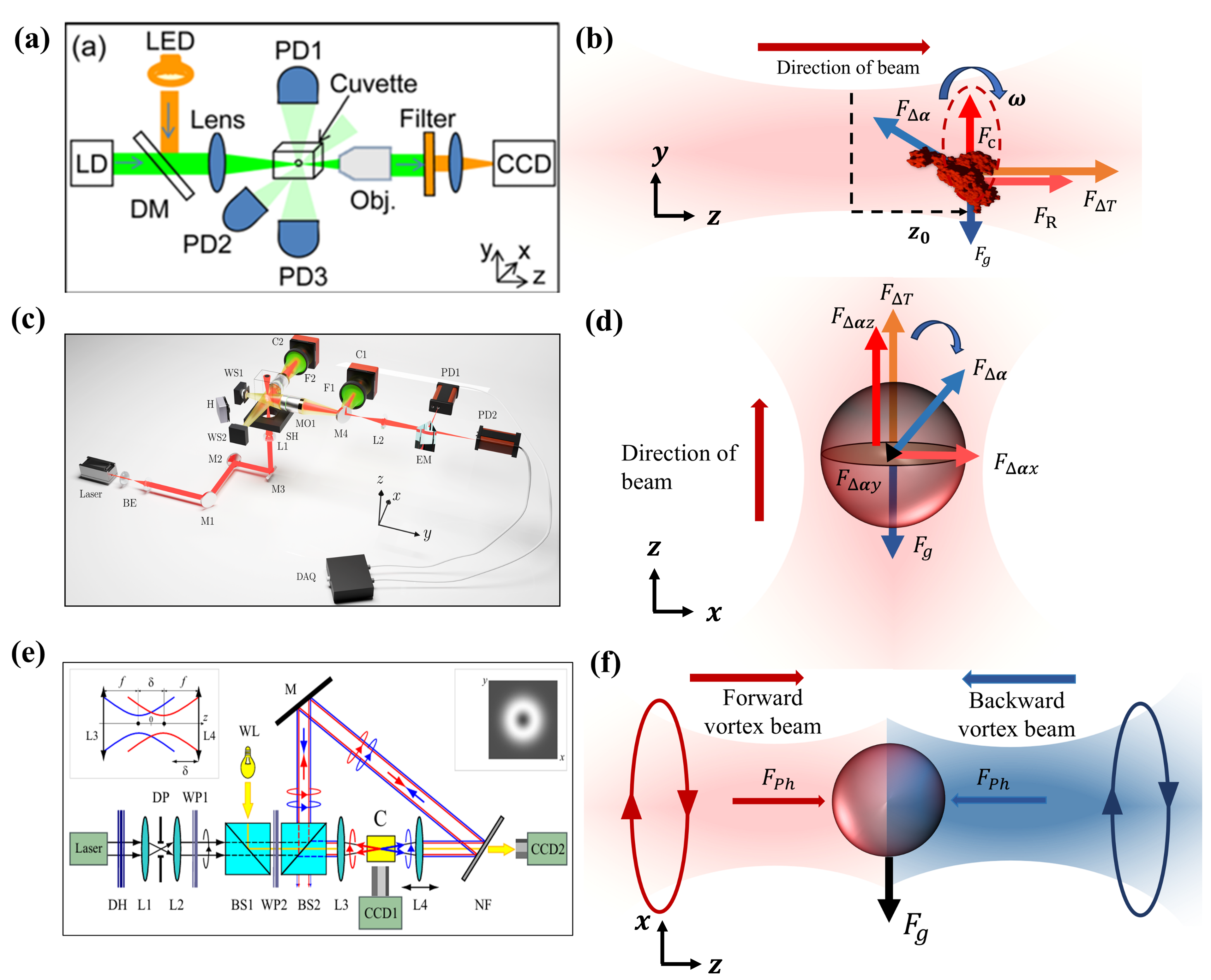}
    \caption{Experimental configurations and corresponding intuitive mechanisms for photophoretic trapping in air: (a, b) horizontal trapping arrangement with its underlying mechanism, where $F_C$ denotes the centripetal force acting on the particle \cite{lin2014optical} \copyright\ 2014 AIP Publishing LLC; (c, d) vertical trapping configuration and the associated trapping process \cite{pahi2025light} \copyright\ 2025 IOP Publishing Ltd, and (e, f) counter-propagating dual-vortex beam setup together with the proposed mechanism. In this case, $F_{ph}$ represents the photophoretic force exerted by the forward and backward vortex beams (indicated by the red and blue arrows, respectively), while the circular rings illustrate the sense of rotation of the forward and backward vortex beams (red and blue, respectively) \cite{shvedov2009optical} \copyright \ 2009 Optica Publishing Group.}
    \label{trap_mech}
\end{figure*}

Considering the trapping mechanisms in vertical and horizontal single-beam configurations, it becomes evident that although such schemes are relatively simple, requiring fewer optical components and offering cost-effectiveness, they often suffer from limited robustness and lower efficiency in stable confinement. To overcome these limitations, counter-propagating beam configurations are employed, where two focused beams are symmetrically aligned to balance the photophoretic forces, offering significantly enhanced stability and robustness \cite{gong2016optical}. This geometry allows stable confinement of absorbing particles using various beam profiles, including Gaussian, hollow, and vortex beams. Vortex beams in a dual-beam configuration have been employed to achieve three-dimensional trapping of aerosol particles, as demonstrated by Shvedov et al. \cite{shvedov2009optical}. In their system, the axial confinement was achieved through the balance of photophoretic forces exerted from opposite sides of the particle, while transverse confinement originated from the bright intensity ring of the vortex beam, which simultaneously compensated for gravity in the vertical plane (shown in Figure \ref{trap_mech}(f)). A crucial aspect of the dual-vortex configuration was the control of the relative direction of rotation, as any mismatch could introduce transverse intensity minima, thereby allowing particles to escape. This issue was resolved by reflecting a single vortex beam ($l=1$) an odd number of times before counter-propagation, ensuring that the transverse intensity distribution remained radially symmetric (experimental setup shown in Fig.~\ref{trap_mech}(e)). Such a configuration not only stabilised the trap but also effectively doubled the orbital angular momentum.  
Beyond vortex beams, stable confinement of particles in counter-propagating geometries has also been experimentally achieved using Gaussian beams, where the beam was reflected and reintroduced into the trapping cell via mirrors \cite{wang2015optical}. Similarly, hollow beams generated with axicons have been counter-propagated to trap diverse microparticles, including grass smut spores and riboflavin \cite{pan2012photophoretic, wang2014experimental, wang2015photophoretic}.  

\section*{Dependence of Photophoretic Trapping on Physical Parameters}

By identifying and tuning the parameters that determine the photophoretic force, controlled manipulation of particles in experiments becomes achievable.

\begin{figure*}[t]
    \centering
    \includegraphics[width=0.9\textwidth]{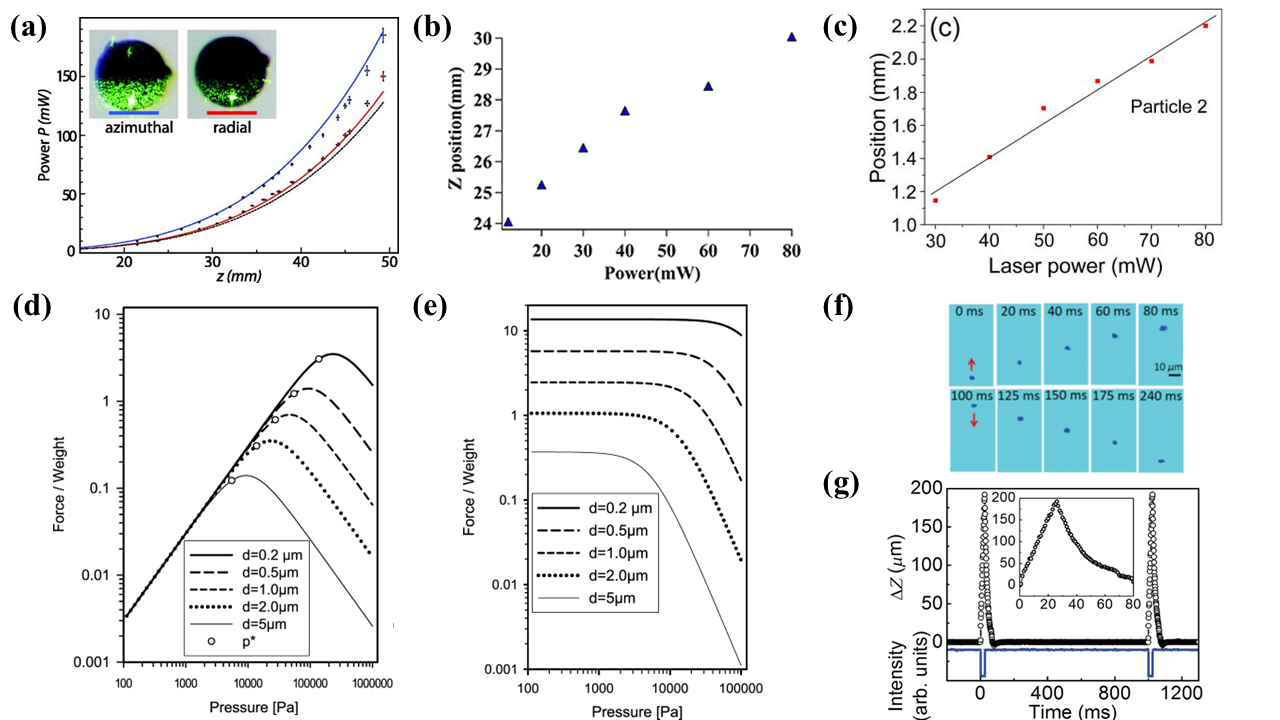}
    \caption{Parameter Dependence of the Photophoretic Force (a) Polarisation dependence of photophoretic force\cite{shvedov2012polarization} \copyright \ 2012, AIP Publishing LLC (b) Power vs position along beam direction in a vertical trapping configuration\cite{pahi2024study} (c) Power vs position along beam direction in a horizontal trapping configuration\cite{lin2014optical} \copyright \ 2014, AIP Publishing LLC  (d) Variation of $\Delta T$ photophoretic force with ambient pressure for particles of different diameters\cite{horvath2014photophoresis} \copyright \ 2014 Hosokawa Powder Technology Foundation (e) Variation of $\Delta \alpha$ photophoretic force with ambient pressure for particles of different diameters\cite{horvath2014photophoresis}. \copyright \ 2014 Hosokawa Powder Technology Foundation (f,g) shows temporal dependence of photophoretic force\cite{chen2018temporal}. \copyright \ 2018, American Physical Society}
    \label{phys_parameters}
\end{figure*}

\subsection*{Polarization}

As we know, trapping in air using photophoretic forces arises from the interaction of light with an absorbing particle, with absorption of light inside the particle playing a crucial role. The absorption of light inside a material depends both on the properties of the particle (such as orientation and anisotropy) and on the properties of the incident light (polarization and intensity). When light interacts with a surface, the transmitted and reflected electric fields differ for different polarizations. Also the particle’s orientation further modifies the relative amounts of transmitted and reflected fields, thereby altering the absorbed power (Eq.~1 of Ref.~\cite{shvedov2012polarization}). Since the Fresnel reflection and transmission coefficients depend on both polarization and angle of incidence, the absorption and consequently the magnitude of the photophoretic force varies accordingly. The balance between gravitational and photophoretic forces determines the trapping position (Eqs.~\ref{horizontal-trap}, \ref{vertical-trap}). Thus, polarization provides an additional degree of control over absorbing particles in a photophoretic trap. This effect has been experimentally demonstrated for large (\(80\text{--}110 \,\mu\text{m}\)) carbon-coated hollow spheres confined in a vertical vortex-beam trap, where switching the polarization from radial to azimuthal, while keeping the laser power constant, resulted in different trapping efficiencies \cite{shvedov2012polarization} also shown in figure \ref{phys_parameters}(a). However, the influence of polarization in single Gaussian beam photophoretic traps, both in modifying trapping efficacy and in shaping particle dynamics, remains largely unexplored. In particular, polarization dependent variations in the dipole (scattering) force, which can be significant for the particle sizes typically used in photophoretic trapping, may shift the equilibrium position and alter the trap stiffness. A systematic investigation of these effects could be an interesting area of research.

\subsection*{Power}

From theory~\cite{desyatnikov2009photophoretic, rohatschek1995semi}, the photophoretic force is directly proportional to the intensity of the illuminating light beam. The light intensity can be expressed as
\[
I = \frac{P}{A},
\]
where \(P\) is the optical power of the beam and \(A\) is the cross-sectional area of the beam at the particle location. In a typical trapping configuration, a balance is achieved between the downward gravitational force and the upward photophoretic force acting on the particle. Thus, for a given particle and ambient condition, a specific light intensity corresponds to the equilibrium trapping position.

In a vertical trapping geometry (Fig.~\ref{trap_mech}(c,d)), when the laser power is reduced, the particle tends to move closer to the beam focus as shown in figure \ref{phys_parameters}(b), where the beam cross-sectional area is smaller and the local intensity is higher ~\cite{pahi2024study,chen2018temporal}. This behavior indicates that the particle continuously adjusts its position to maintain a constant local intensity, thereby preserving force equilibrium. Conversely, as the laser power increases, the particle moves away from the focus toward regions of larger beam cross-section, where the intensity again matches the value required for equilibrium. A similar dependence on laser power is observed in horizontal trapping configurations as well~\cite{lin2014optical} as shown in figure \ref{phys_parameters}(c), confirming that the trapping position is governed by the local light intensity rather than the absolute laser power. Consequently, by varying the beam power, the particle can be translated along the beam axis, effectively enabling controlled axial manipulation in a photophoretic trap.

\subsection*{Pressure}

The magnitude and direction of the photophoretic force acting on illuminated aerosol particles depends strongly on the surrounding gas pressure. This behavior originates from the way gas molecules exchange momentum and energy with the particle surface, which changes with the mean free path of the gas relative to the particle size. At \emph{low pressures} (free-molecule regime, high Knudsen number), gas molecules interact less frequently with each other, and the photophoretic force increases approximately \emph{linearly with pressure} because a larger number of gas surface collisions enhances the net momentum transfer. In contrast, at \emph{high pressures} (continuum regime, low Knudsen number),
the increased frequency of molecular collisions more efficiently dissipates the momentum and temperature gradients in the gas around the particle, leading to a reduction in the net force. In this limit, the photophoretic force becomes \emph{inversely proportional to pressure}. Between these two regimes, the force reaches a \emph{maximum at an intermediate pressure}, where the mean free path is comparable to the particle size. This characteristic non-monotonic trend, a rise, a peak, and then a decay with increasing pressure is a fundamental feature of (\(\Delta T\)) photophoretic force \cite{horvath2014photophoresis}as shown in figure \ref{phys_parameters}(d). But, the magnitude of photophoretic (\(\Delta \alpha\)) force remains constant at low pressure and tends to decrease as pressure increases (figure \ref{phys_parameters}(e)). The detailed semi empirical treatment of pressure dependence of photophoretic force can be found in Rohatschek's work\cite{rohatschek1995semi}.  Interestingly, Rohatschek's framework remains the most widely referenced description of pressure-dependence of photophoretic force.

\subsection*{Temporal Dependence}

To achieve precise control in optical manipulation using photophoretic forces, understanding their temporal response is essential. Whereas radiation pressure arising from the direct momentum transfer from light to the particle acts instantaneously, photophoretic forces that originate from an indirect interaction mediated by gas molecules does not act instantaneously. Because the particle requires a finite time to heat up after light absorption, the resulting photophoretic force does not respond immediately to changes in light intensity. Consequently, a temporal lag emerges between the modulation of optical intensity and the force experienced by the particle. 

This temporal behavior was experimentally investigated by Chen \textit{et al.}~\cite{chen2018temporal} using strongly absorbing carbon-based microspheres trapped in air by a Gaussian beam using negative photophoresis. The force balance along the beam propagation direction for their case was 
\begin{equation}
    F_{\mathrm{pp}} = F_{G} + F_{\mathrm{RP}},
\end{equation}
where $F_{\mathrm{pp}}$ is the photophoretic force, $F_{G}$ is the gravitational force, and $F_{\mathrm{RP}}$ is the radiation-pressure force. The laser power was modulated with a square pulse. When the laser power was switched off, $F_{\mathrm{RP}}$ dropped to zero instantaneously. If the photophoretic force had also vanished immediately, the particle would have begun falling due to gravity; however, the experiment showed that the particle initially moved upward for a short duration before eventually descending. This counterintuitive behavior, visible in Fig.~\ref{phys_parameters}(f,g), clearly demonstrates the time-delayed response of the photophoretic force.

A theoretical model incorporating a finite thermal response time successfully reproduced the observed particle trajectories. The characteristic time constant for the particles used in the experiment was found to be on the order of $0.1$~s, confirming that photophoretic forces evolve on timescales much slower than the instantaneous action of radiation pressure.

\section*{Trapping with Various Beam Profiles}
In the previous section, we discuss different experimental configurations for trapping several kinds of particles employing photophoretic forces. It has been observed that the geometry or profile of a beam plays a crucial role in photophoretic trapping. The fundamental question that triggers the photophoretic trapping community is whether beams with complex beam profiles would lead to stronger photophoretic traps. It is also worth noting that investigating the physics behind the improvement of trapping efficiency for complex beams is a worthwhile endeavor. Therefore, in this section, we provide an overview of various beam profiles that have been implemented for photophoretic trapping, focusing on their trap robustness, applications, and the underlying physics. \\

\begin{figure*}[!ht]
\centering
\includegraphics[width=0.9\textwidth]{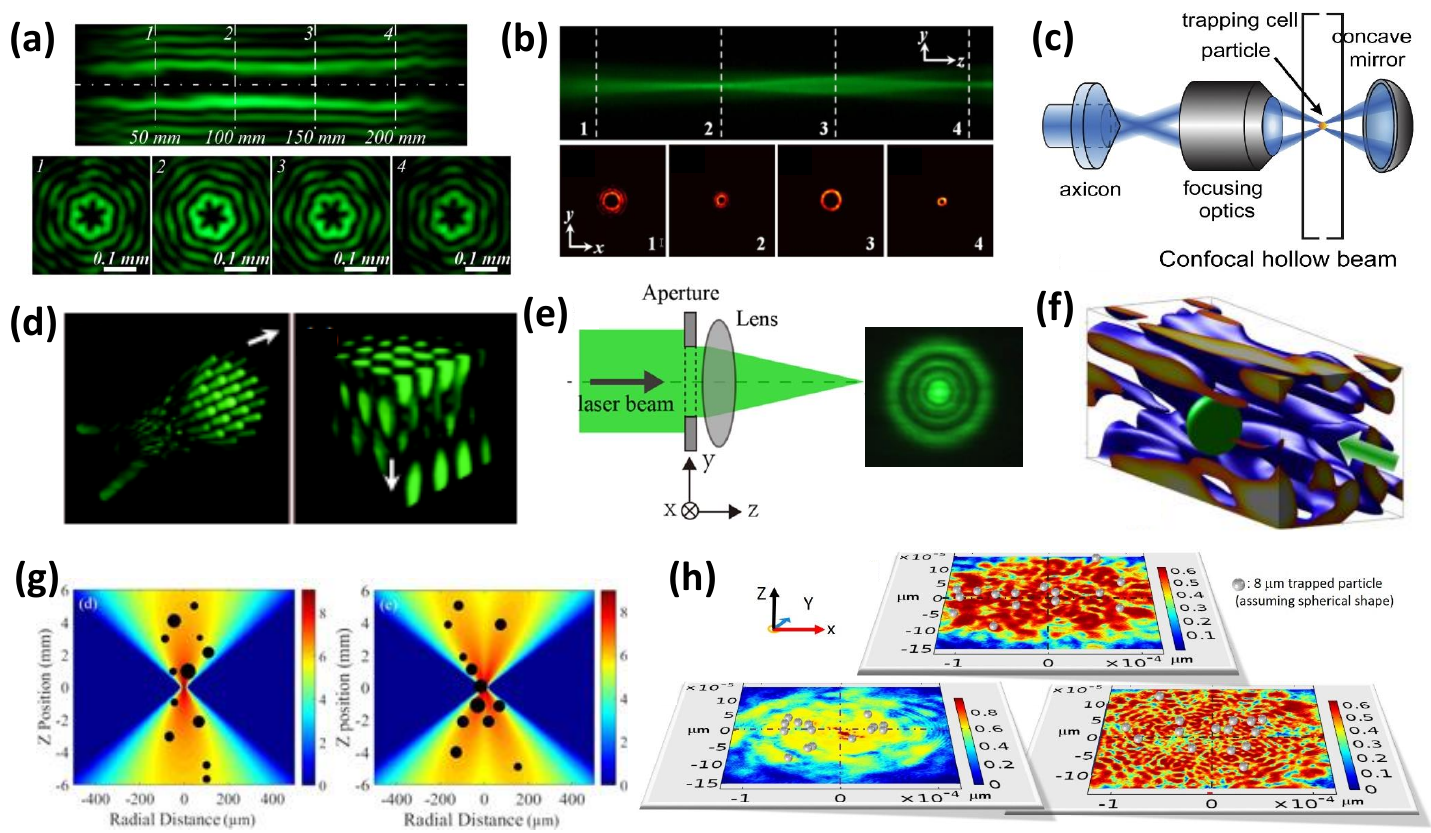}
\caption{ (a) Intensity distribution of dark-hollow beam, made by superimposing two Bessel beams \cite{porfirev2015dark}  \copyright\ 2015 Optical Society of America. (b) Experimental photography of a single optical bottle beam, where the two bottle necks are located near the planes 2 and 4. At the bottom, snapshots of the transverse intensity patterns of the bottle beam (contrast-enhanced) taken at planes 1–4 are marked. \cite{zhang2011trapping} \copyright \ 2011 Optical Society of America; (c) Experimental configuration of generating a confocal-beam trap using a hollow beam, which was generated through axicons. \cite{gong2016optical} \copyright \ 2016 AIP Publishing LLC. (d) Optical lattice structure profile for trapping multiple particles. \cite{shvedov2012optical}  (e) Left: Generation of taper-ring optical field using aperture and lens. Right: Taper-ring light distribution of one transverse section. \cite{liu2013manipulation} \copyright \ 2014 Optical Society of America   (f) Numerical 3D modelling of the speckle intensity distribution for a small volume 5 $\mu m$ $\times$5 $\mu m$ $\times$ 15 $\mu m$ inside the trapping region with an average speckle size of 2 $\mu m$ in diameter. \cite{shvedov2010laser} \copyright \ 2010 IOP Publishing Ltd (g) Trapping locations of the trapped absorbing particles in Gaussian (left) and superposition of Gaussian + Hermite-Gaussian (HG) (right) beams after determining the radial and axial positions of the trapped particles. Particles are not drawn to scale, though their size ratio has been maintained.\cite{sil2017dual} \copyright \ 2017 IOP Publishing Ltd  (h) Normalized trapped particle position distribution over a transverse plane of the specific beam size for three different speckle patterns generated from a multi-mode fiber. \cite{sil2024ultrastable} \copyright \ 2024 American Chemical Society}
\label{structure_profile}
\end{figure*}

In 1982, a 20 $\mu m$ diameter dye-impregnated glycerol sphere was first levitated in air at 30 Torr atmospheric pressure using a doughnut beam of Ar+ laser employing the rediometric pressure force \cite{lewittes1982radiometric}. Then, multiple absorbing metal-oxide particles were optically levitated using a distorted Gaussian beam based on the thermal creep phenomenon \cite{huisken2002optical}. But the idea of using a structured beam was first provided by Shevdov
et al., who demonstrated this by publishing a series of research articles between 2009-2012. Thus, in 2009, Shredov et al. first demonstrated stable positioning, guiding, and trapping of multiple absorbing carbon nanoclusters of average individual nanoparticle size around 6 nm in air using two counter-propagating and co-rotating optical vortex beams, generated by a diffraction fork-type hologram. Further, a long-range transport of  $\sim$ 100 $\mu m$ particles over a meter scale was performed by using a vortex beam generated from a diffraction hologram \cite{shvedov2010giant}. Additionally, optical hollow-cone beams generated using two axicons \cite{redding2015optical, redding2015photophoretic} or even overlapping Bessel beams generated from a spatial light modulator \cite{porfirev2015dark} (see Fig.\ref{structure_profile},(a)), have been utilized for photophoretic trapping and micromanipulation, as well as for chemical and physical analysis of absorbing particles, including various bioaerosol particles \cite{gong2018optical,lamhot2010self}. Recently, an experimental technique was reported where a Bessel-like beam was generated by utilizing the combination of single mode fiber (SMF) and multimode fiber (MMF) and subsequently focused by a glass sphere lens, which could be used for trapping of multiple absorbing particles in a liquid medium \cite{zhang2025trapping}. This configuration can be extended for trapping absorbing particles in air medium. Moreover, to enhance the trapping flexibility, holographic beam shaping was used to generate optical bottle beams either by a Moiré technique \cite{zhang2011trapping} (see Fig.\ref{structure_profile} (b)), or by convolving the trapping geometry with discrete trapping sets, creating a zero-intensity region surrounded by light in all three dimensions  \cite{alpmann2012holographic}. In addition, a bottle beam can be generated utilizing the spherical aberration - a unique property of a lens with two spherical surfaces \cite{shvedov2011robust}, or from the conical refraction of light in a biaxial crystal \cite{turpin2013optical,esseling2018conical} for trapping absorbing droplets. Similarly, to improve the trapping and manipulation of multiple particles in gaseous media, two methods were reported. One was a three-dimensional (3-D) optical lattice generated by overlapping coherent beams, arising from a two-dimensional (2D) amplitude diffraction grating \cite{shvedov2012optical} as shown in Fig \ref{structure_profile} (d), while the other was a tapered-ring optical field \cite{liu2013manipulation, liu2014photophoretic} generated by diffusing a laser beam from a circular aperture, followed by weakly focusing it (see Fig.\ref{structure_profile} (e). It was reported that using this field pattern with moderate laser power (beyond 100 mW), several hundred irregularly absorbing particles could be captured and revolved around the optical axis in different planes. Furthermore, to simplify the complexity, such structured light fields can be created using a speckle-pattern - a high-contrast, fine-scale granular pattern - generated from the interference of a large number of dephased but coherent monochromatic waves \cite{goodman1976some}. 

On another note, speckle optical tweezers (STs) have been developed, utilizing a speckle pattern generated by a multimode fiber to collectively manipulate high-refractive-index particles \cite{volpe2014brownian}  and control the motion of both high- and low-refractive-index microparticles, as well as nanoparticle-loaded vesicles. However, STs have predominantly been employed in liquid media and demonstrated trapping in two dimensions \cite{jamali2021speckle}. For photophoretic-based trapping, Shvedov et al. demonstrated that a volume speckle field (see Fig.\ref{structure_profile} (f)) generated by a coherent laser beam and a diffuser can be used to confine a large number of carbon particles in air, employing photophoretic forces \cite{shvedov2010selective, shvedov2010laser}. Recently, we experimentally demonstrated a robust fiber-based trapping system using a single multi-mode (MM) fiber, where a speckle pattern from the MM fiber facilitated significantly more robust optical traps for trapping and manipulation of mesoscopic particles in all three-dimensions compared to Gaussian beam profile \cite{sil2024ultrastable}. \\

However, besides this unidirectional beam configuration for trapping, structured beams, such as a vortex beam \cite{shvedov2009optical} or a hollow beam, were used to create a counter-propagating beams trapping configuration, which enhances the trapping stability and efficiency. For example, two counter-propagating hollow beams were focused by two long-working-distance objectives to form 3D cones, allowing particles to be trapped, rotated, and bounced between the edges of the cones for use in spectroscopy applications \cite{pan2012photophoretic,wang2014experimental,wang2015photophoretic}. Although dual-beam traps provide high trapping stability, they require two trapping beams, which further increases the cost, occupies more space, and requires additional precision alignment. To mitigate this, the confocal-beam trap was introduced, which enhanced the robustness of dual-beam traps. A confocal-beam trap can be constructed using a focusing lens and a spherical concave mirror where the lens focuses a collimated beam, and the concave mirror reflects the focused beam in the reverse direction - creating a virtual counter-propagating optical field as shown in Fig.\ref{structure_profile} (c). Gong et al. developed a confocal-beam trap using a Gaussian beam and a hollow beam for trapping a single nanotube cluster with higher stability \cite{gong2016optical}. Also, it was demonstrated that the hollow optical field could be able to trap multiple particles and control their number by tuning the distance between the focusing lens and the concave mirror or even using external perturbation \cite{gong2016optical, gong2017characterization}. Later, based on the confocal beam configuration with a hollow beam, a universal optical trapping (UOT) system evolved, which was used to study the chemical or physical properties of aerosols or airborne droplets through time- and position-resolved Raman spectroscopy of single chemical aerosol droplets \cite{kalume2017detection,kalume2017position}. 

From the above descriptions, it is clear that the general understanding, as is clear from the various experimental configurations and studies, that particles required a dark region to be trapped, with a bright region surrounding it. This led to a strong conception that photophoretic trapping occurs only in the dark regions of the light field. However, this is not entirely accurate, as our recent work, in collaboration with studies from other groups \cite{sil2020study, bera2016simultaneous, sil2017dual, sil2022trapping,lin2014optical, he2019investigation, pahi2023comparison}, has shown that particles can also be trapped by photophoretic forces associated with a fundamental Gaussian beam, indicating that trapping does not necessarily require particles to reside in a dark region of the beam.\\

We now move on to the quantification of the increase of the photophoretic forces using the structured beam profiles. In the section on the principles of trapping using photophoretic forces, where we discussed the trapping mechanism, we noted  that for the vertical configuration, particles are confined in the axial direction by a balance of the longitudinal component of the photophoretic force and gravity, so that the process can be described as optical levitation. Along the radial direction, however, the transverse photophoretic body force applies a torque on particles due to its interaction with gravity \cite{bera2016simultaneous}, leading to a possible helical motion. Perhaps this was the reason why the trapped particles were observed to be radially shifted off-axis with respect to the trapping beam center. In addition, the trap stiffness was seen to be linearly proportional to the intensity. Based on this concept, Sil et al. generated a structured beam by superimposing the Gaussian mode and the first-order Hermite-Gaussian (HG) mode, which was generated from a single-mode fiber that supports dual-modes at the operating wavelength of the laser. The presence of the HG mode ensured that the intensity maximum was shifted off-axis radially. A simulated propagation of the Gaussian beam and the superposition of Gaussian and first-order HG beam is shown in Fig.\ref{structure_profile} (g), where the exact location of the trapped paritcles were depicted by experimentally measuring the trapped particle coordinate. It was observed that the average intensity sampled by particles in the Gaussian + HG beam was greater than that in the Gaussian beam by a factor of around 1.9. Therefore, this 3-D optical trap using a single commercial fiber demonstrated that the superposition mode was more effective in trapping and manipulation compared to the fundamental mode by around 80 \% \cite{sil2017dual}.\\

In continuation of this work, Sil et al. further demonstrated an ultra-stable, robust fiber-based trap system by introducing a multi-mode fiber instead of a single-mode fiber \cite{sil2024ultrastable}. As described earlier, a speckle pattern generated from the MM fiber was further focused using a long-focal-length lens, which proved effective for trapping and manipulation. Note that the speckle pattern produced from an MM fiber has the advantages of uniform distribution, easy alignment, high transmission efficiency, and flexibility. It was found that the MM profile provided a transverse trapping force approximately eight times stronger than that of a single-mode Gaussian beam, enabling high manipulation speeds of roughly 5 mm/s in both axial and transverse directions.  This trapping system was capable of confining a single micron-sized particle for 6-7 hours or longer, all without the need for a high-NA microscope objective \cite{sil2024ultrastable}. In addition, it was observed that the photophoretic trapping force is maximum when the average size is comparable to the particle size. Fig.\ref{structure_profile} (h) shows three different modes based on the speckle size, obtained by changing the coupling angle of the input beam, along with the corresponding experimentally measured normalized particle-position distributions for a given beam size.  It was reported that particles could remain in the bright region of the beam by exhibiting a complex dynamical motion. Thus, a single MM fiber-based photophoretic trap system offers significant advantages in terms of robustness, portability, ease of use, and the facilitation of diverse applications towards simultaneous trapping, micromanipulation, and spectroscopy of
aerosols/bioaerosols.  

\section*{Dynamics of Particles in a Photophoretic Trap}
The particles confined in a photophoretic trap in air exhibit a plethora of interesting dynamical behaviours depending on the trapping configuration and particle properties. These behaviours range from circular or elliptical orbits - occurring either as rotation around the particle’s body axis or as orbital motion about a particular axis \cite{pahi2024study, horvath2014photophoresis, lin2014optical} to vigorous, enhanced Brownian-like motion \cite{pahi2025light}, as well as complex trajectories that combine multiple types of motion \cite{horvath2014photophoresis}. A systematic characterisation of these diverse dynamical regimes requires the application of statistical tools. In conventional optical tweezers systems, the characterisation of particle dynamics and trap parameters typically relies on statistical tools, which include the computation of the autocorrelation function to reveal temporal correlations, power spectral density (PSD) analysis to identify characteristic frequencies of motion, mean-squared displacement (MSD) analysis to capture diffusive or ballistic behaviour, probability distribution functions (PDFs), velocity–autocorrelation functions (VACF) and so on\cite{gieseler2021optical}. Extending the same framework to photophoretic trapping, where the forces are often non-conservative and the dynamics more diverse, allows one to systematically characterise the trap stiffness, stability, effective potential landscape, and non-equilibrium features of trapped particles. Most importantly, the characterisation of trap stiffness is of central importance as it quantifies the restoring force experienced by a trapped particle and its dependence on parameters such as laser power. In optical tweezers, stiffness is typically evaluated by fitting the PSD or ACF of particle fluctuations, or applying the equipartition theorem for calibration \cite{gieseler2021optical}. In contrast, within the photophoretic trapping community, researchers have employed certain experimental approaches to determine the stiffness of the trap. In the following, we review the various methods reported in the literature for stiffness characterisation in photophoretic trapping systems.

\subsection*{Trap stiffness and stability
}
Several experimental techniques have been developed to quantify the stiffness of photophoretic traps in both horizontal and vertical configurations, as well as to study its dependence on optical power. One technique reported in the literature involves modulating the beam intensity using an acousto-optic modulator (AOM) and recording the particle’s response across different modulation frequencies. The system was modelled as a spring–mass oscillator incorporating a time-dependent sinusoidal stiffness term to represent the modulation in optical intensity, and the resulting response curve was fitted to extract the trap stiffness \cite{lin2017measurement}. However, intensity modulation inherently alters the levitation conditions, leading to shifts in the equilibrium position of the trapped particle and introducing challenges in accurately tracking its motion. A more refined approach investigated trap stiffness through motional resonances induced by spatially modulating the trapping beam. In this method, the amplitude and phase response of the trapped particle under external modulation across a range of frequencies were fitted to a forced-damped oscillator model to determine the trap stiffness. This technique has been implemented using a multi-sine driving signal, wherein a superposition of sinusoidal modulations at different frequencies was applied and the corresponding particle response analysed. The key advantage of this method is that it enables rapid frequency scanning compared to the single-sine approach, which requires separate measurements at each frequency. Sil et al. \cite{sil2020study} demonstrated this method for both single-sine and multi-sine modulations, where the beam was spatially modulated using a piezo-actuator. A further notable technique was introduced by Zhu \textit{et al.} \cite{zhu2024measurement}, who employed two hollow counter-propagating beams to create spatially separated trapping sites. When one of the beams was switched off, the trapped particle transitioned to the adjacent trap, and the recorded flight trajectory was fitted with a Langevin model (neglecting the inertial term) to evaluate the trap stiffness.\\

\par
\begin{figure}[h]
    \centering
    \includegraphics[width=\linewidth]{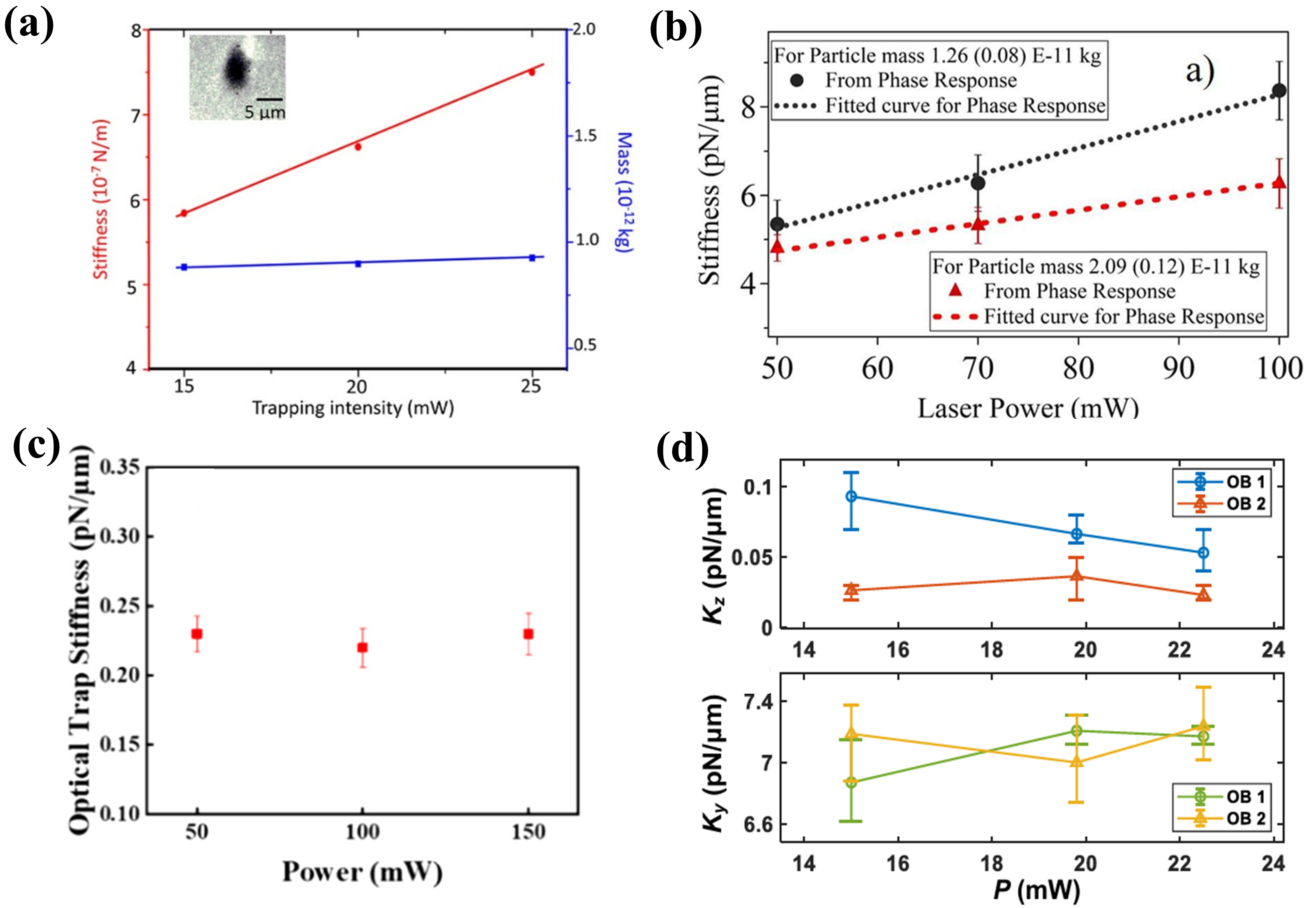}
    \caption{Variation of trap stiffness with laser power: (a) intensity modulation using an AOM (red curve shows stiffness as a function of trapping intensity)\cite{lin2017measurement} \copyright\ 2017 Optical Society of America, (b) spatial modulation via a multi-sine signal (red and black lines correspond to particles of different masses) \cite{sil2020study} \copyright\ 2020 AIP Publishing LLC,  (c) flight-based method, where the particle transitions between adjacent traps when one of the counter-propagating beams is switched off \cite{zhu2024measurement} \copyright\ 2024 AIP Publishing LLC, and (d) trapping of absorbing particles in an optical bottle (OB) beam generated by dynamic holography (with axial and transverse stiffness versus power shown in the top and bottom panels for two OBs, respectively) \cite{xu2025dynamic} \copyright \ 2025 Optica Publishing Group \& Chinese Laser Press. While (a) and (b) show an increasing stiffness with power, both (c) and (d) exhibit stiffness that remains independent of laser power.}
    \label{stifness_vs_power}
\end{figure}

 
Interestingly, divergent observations have been reported regarding the variation of trap stiffness with laser power. In the case of spatial modulation (via multi-sine signal) or intensity modulation (via an AOM), an increase in trap stiffness with optical power has been observed for irregular carbon toner particles as well as clusters of CuO powder \cite{sil2020study,lin2017measurement} (shown in Figure \ref{stifness_vs_power}(a, b)). This behaviour was attributed to a shift of the equilibrium position with increasing laser power, such that the particle resided at locations where the photophoretic force counterbalanced gravity. In this process, the accommodation coefficient is believed to be modified by the local laser intensity, thereby altering the strength of the photophoretic force and leading to an enhancement in trap stiffness. In contrast, using the flight-based method in a counter-propagating beam configuration, Zhu \textit{et al.} \cite{zhu2024measurement} reported that the trap stiffness remains nearly independent of laser power (Figure \ref{stifness_vs_power}(c)). Here, the local intensity distribution and the corresponding photophoretic force at the trapping sites scaled proportionally with power, resulting in a stiffness that remained effectively unchanged. A similar power-independent behaviour of trap stiffness has also been reported by Xu et al. \cite{xu2025dynamic} for absorbing particles trapped (Figure \ref{stifness_vs_power}(d)), released, and transported in an optical bottle (OB) beam using a dynamic holographic approach using a spatial light modulator (SLM), where confinement in a three-dimensional dark region prevented continuous laser illumination, resulting in a trap stiffness (obtained by fitting PSD with Lorentzian function) that showed no clear correlation with optical power.

\subsection*{Observation of diverse dynamic behaviour}
Although the quantification of trap stiffness provides important insights into the restoring forces in photophoretic trapping, the use of irregularly shaped particles introduces the additional complexity of an anisotropic morphology, which can lead to coupling of different degrees of freedom, resulting in very different stiffness values. This complexity highlights the fact that that stiffness alone cannot fully characterise the trapping potential or the rich non-equilibrium features of the system. A deeper understanding emerges only through the study of particle dynamics, which encodes the subtle interplay between photophoretic forces, gravity, and stochastic forces. In this context, Pahi et al. \cite{pahi2025light, pahi2024study} identified three characteristic dynamical regimes in the vertical configuration which arises spontaneously: (i) rotation about the particle’s body axis, called as spinning clusters, (ii) coupled spin–orbital motion with the orbit direction governed by the cluster’s morphology, called as orbiting clusters and (iii) a non-rotating state, in which the particle remains fixed in angular orientation called as non-rotating clusters. In all three cases, the particle underwent motion along the beam direction. In the first two cases, this motion manifested as periodic oscillations, whereas, in the third case, the particle exhibited intermittent axial displacements (“kicks”) that occur with irregular intervals. Note however, that no quantitative theoretical model still exists which would explain such intriguing dynamics as emergent behaviour from an interplay of interesting physical phenomena concerning photophoretic forces. 
\par 
We now turn to the key experimental observations associated with these different dynamical regimes. For spinning clusters, the rotation frequency increased with laser power but eventually saturated at an approximately constant value, while the amplitude of axial oscillations decreases correspondingly, as shown in Fig.~\ref{Rot_1}(a). In contrast, for orbiting clusters, the rotation frequency decreased with increasing laser power, accompanied by an increase in both the orbital diameter and the axial oscillation amplitude, as shown in Fig.~\ref{Rot_1}(d). Thus, in both spinning and orbiting regimes, the axial oscillation amplitude were anticorrelated with the spinning frequency. This interplay provided a means to tune one type of motion by adjusting the other. Turning to the non-rotating regime, observations revealed that particles exhibited directional motion along the axial direction, punctuated by irregular “kick-like” displacements. The amplitude of these axial fluctuations were significantly larger than that of the transverse fluctuations, underscoring a pronounced anisotropy in the dynamics along the axial and transverse directions.\\

To move beyond qualitative trends and achieve a more rigorous description of these regimes, it is essential to employ a theoretical framework, not just to explain but even to simulate the experimental observations. In this context, the Langevin model of a microparticle in harmonic confinement provides a natural starting point for capturing the stochastic and deterministic forces that govern the observed behaviours \cite{pesce2020optical, gieseler2021optical}.


\subsection*{Explaining dynamics from Langevin equation}
\par 
As a cornerstone in the modelling of Brownian dynamics in optical tweezers \cite{gieseler2021optical}, the Langevin equation provides a powerful basis for understanding the diverse dynamical behaviours observed in photophoretic trapping. In what follows, we first introduce the one-dimensional Langevin equation for a spherical particle, and then extend the framework to a two-dimensional form (axial and transverse direction) for an ellipsoidal particle (bringing mobility coupling) and some additional stochastic forces to explain the experimentally observed features of these dynamic regimes\cite{pahi2025light}.
\par 
For a microparticle under harmonic confinement, the one-dimensional form of the equation can be expressed as \cite{li2013brownian}: 
\begin{equation}\label{langevin}
m\ddot{x}(t) + \gamma \dot{x}(t) + kx(t) = \xi(t),
\end{equation}
where \(m\) is the particle mass and the first term, \(m\ddot{x}(t)\), represents the inertial contribution. The second term, \(\gamma \dot{x}(t)\), accounts for viscous damping due to the surrounding medium, while the third term, \(kx(t)\), describes the harmonic restoring force arising from the trap. For a spherical particle of radius \(a\), the drag coefficient is given by Stokes’ law, \(\gamma = 6 \pi \eta a\), where \(\eta\) is the viscosity of the medium. The resonance frequency of the Brownian harmonic oscillator is given by,
\[
\Omega = \sqrt{\frac{k}{m}}.
\] The stochastic force \(\xi(t)\) on the right-hand side accounts for thermal fluctuations and is modelled as a Gaussian-distributed, delta-correlated random process with zero mean and variance determined by the fluctuation–dissipation theorem: 
\begin{equation}\label{noise_prop}
\langle \xi(t) \rangle = 0, \quad 
\langle \xi(t)\xi(t') \rangle = 2 \gamma k_B T \, \delta(t-t'),
\end{equation}
where \(k_B\) is the Boltzmann constant, \(T\) is the ambient temperature, and \(\delta(t-t')\) is the Dirac delta function, ensuring the noise is uncorrelated in time. This correlation of the stochastic force satisfies the ``Fluctuation-Dissipation theorem'', linking the magnitude of thermal fluctuations to the dissipative drag in the system.
\par
\begin{figure*}[h]
    \centering
    \includegraphics[width=\textwidth]{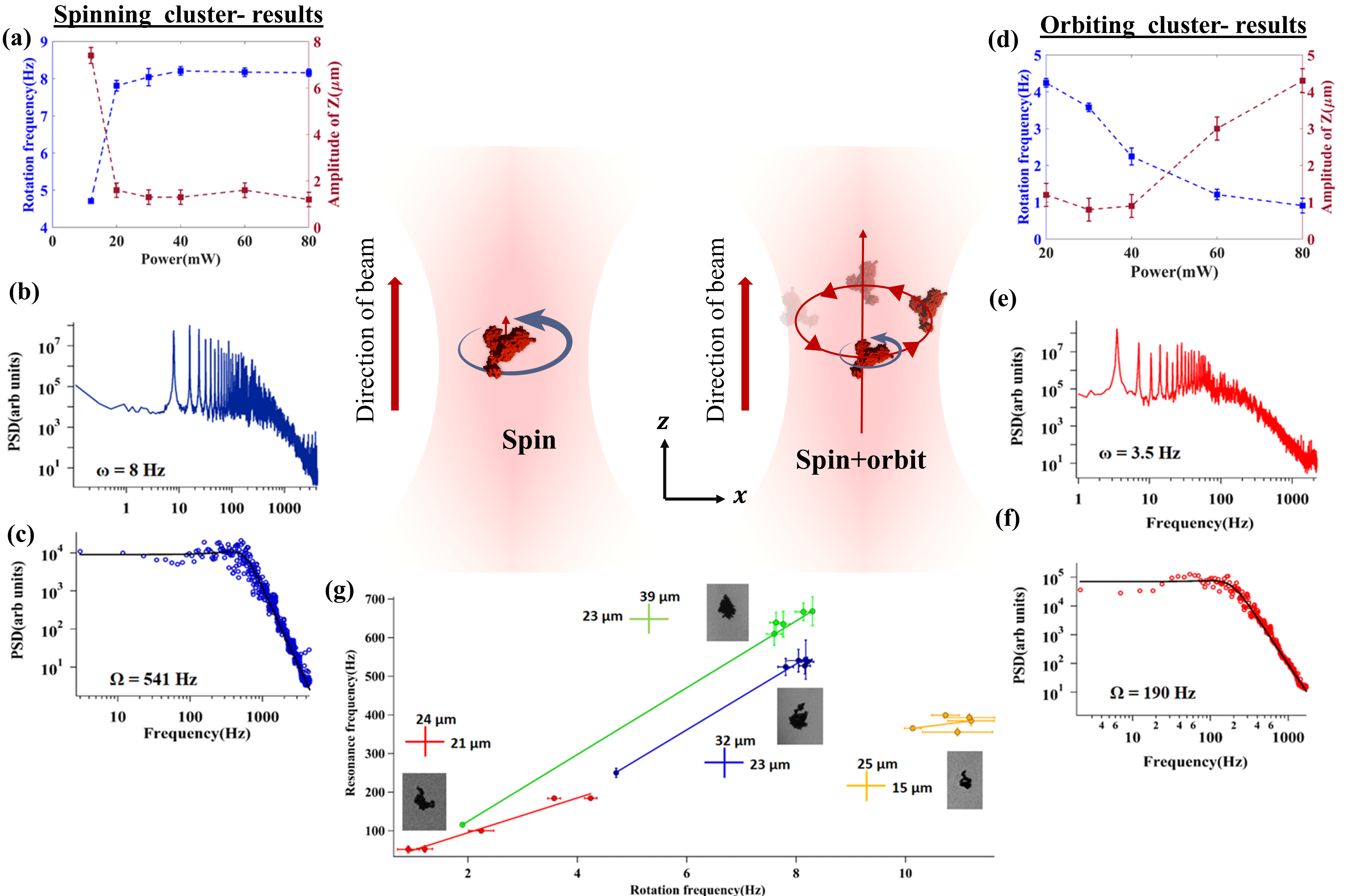}
    \caption{Schematic and experimental results for spinning and orbiting clusters in a photophoretic trap \cite{pahi2024study} \copyright\ 2024 The Author(s). Published by IOP Publishing for IOP and DPG. The central panel illustrates a microcluster confined by a Gaussian beam, exhibiting spinning motion (left) and orbiting motion (right). Panels (a, d) present the dependence of rotation frequency and the amplitude of axial motion on laser power for spinning and orbiting clusters, respectively. Panels (b–c) show the PSD of spinning clusters before and after removal of the rotational peaks, while panels (e–f) display the corresponding PSDs for orbiting clusters. Panel (g) highlights the correlation between the spin and resonance frequencies of the microcluster.}
    \label{Rot_1}
\end{figure*}

 
Now, we extend our discussion to the dynamics of spinning and orbiting microclusters within the framework of a simplified one dimensional Langevin model along the direction transverse to the beam \cite{pahi2024study}. In the presence of rotation --- an inherently two-dimensional motion -- the power spectral density (PSD) measured along the transverse direction (
$x$) exhibits additional rotation-induced peaks, as shown in Fig.~\ref{Rot_1}(b,e). Since the one-dimensional Langevin equation does not incorporate rotational contributions, these features must be removed from the raw PSD to accurately determine the resonance frequency. Accordingly, the PSD was processed to eliminate the rotation peaks, yielding the peak-removed PSD shown in Fig.~\ref{Rot_1}(c,f), which was then fitted using the functional form:


\begin{equation}\label{psd_underdamped}
S_{xx}(\omega) = A \frac{\Omega_x^2 \; g}{(\Omega_x^2 - \omega^2)^2 + \omega^2 g^2},
\end{equation}
where $\Omega_x$ is the resonance frequency of the trap, 
$g = \gamma / m$ is the damping rate, 
$k_B$ is the Boltzmann constant, 
$T$ is the ambient temperature, and 
$\gamma$ is the viscous drag along the observation direction. From this analysis, it was observed that for both spinning and orbiting clusters, there exists a direct correlation between the spin frequency and the trap stiffness, which was inferred from the resonance frequency of the oscillator (Figure \ref{Rot_1}(g)).
\par
\begin{figure*}[htbp]
    \centering
    \includegraphics[width=\textwidth]{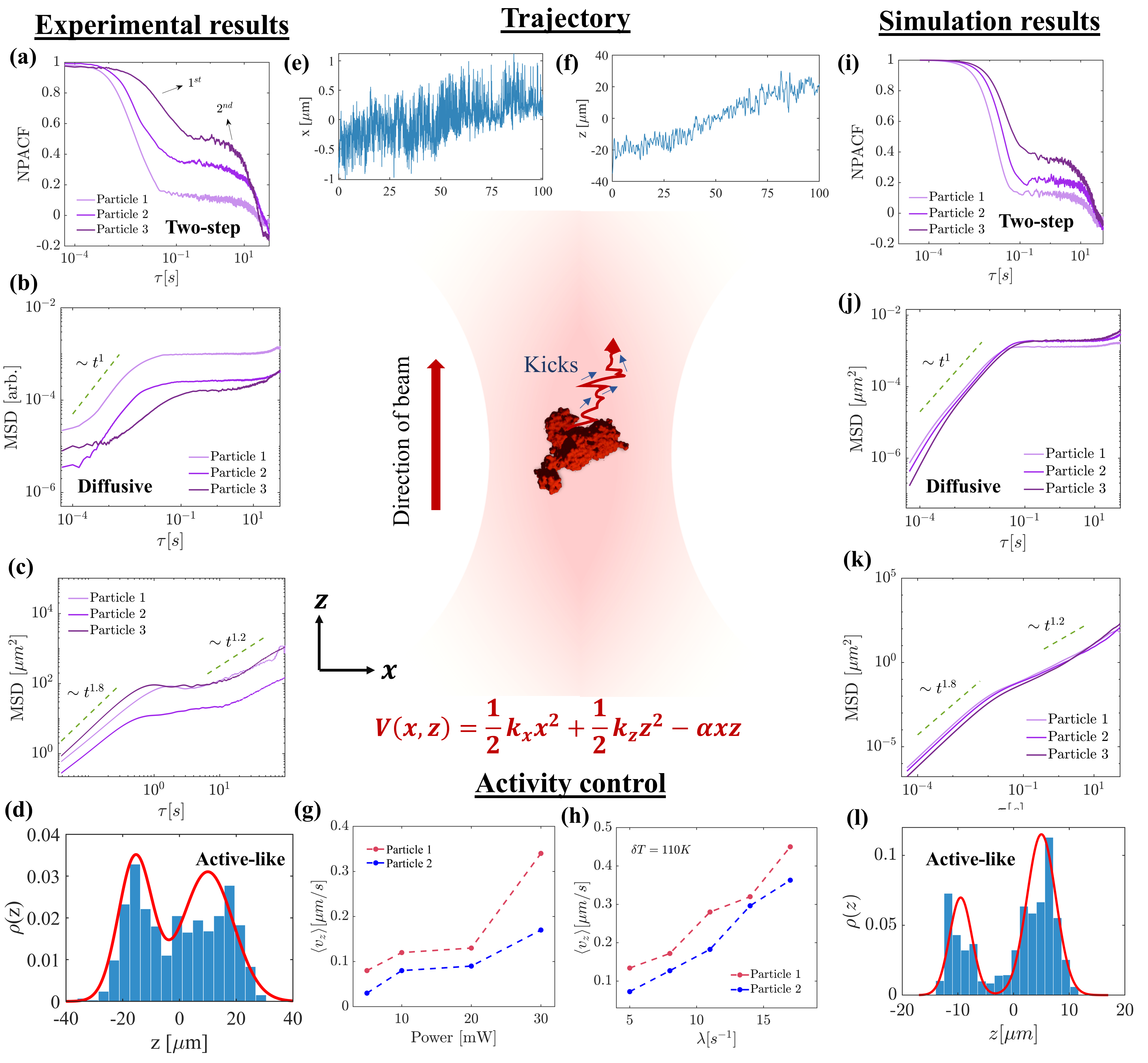}
    \caption{Schematic and experimental/simulation results for non-rotating microclusters in a photophoretic trap \cite{pahi2025light} \copyright\ 2025 IOP Publishing Ltd. The central panel illustrates a microcluster (approximated as an ellipsoid) trapped by a Gaussian beam and subjected to intermittent kicks. Panels (a–d) show the experimental ACF$_x$, MSD$_x$, MSD$_z$, and PDF$_z$, respectively. Panels (e–f) display the corresponding experimental trajectories along the $x$ and $z$ directions. Panels (i–l) present the simulation results reproducing the same quantities. Panels (g–h) demonstrate that the laser power in the experiment corresponds directly to the kick rate in the simulation.}
    \label{non_rot_1}
\end{figure*}

Having understood the characterisation of rotating clusters, we next focus on non-rotating clusters \cite{pahi2025light}. As noted earlier, these systems exhibited intriguing features, most prominently, the sudden kick-like displacements along the beam axis, which added complexity to the dynamics and motivated deeper investigation. Analysis of the autocorrelation function (ACF$_x$) along the transverse direction revealed a characteristic two-step relaxation, whereas the corresponding mean-squared displacement (MSD) exhibited a diffusive behaviour at shorter time scales, an unexpected outcome for a particle trapped in air. In the axial direction, the MSD exhibited two scaling regimes: $t^{1.8}$
 at short times and $t^{1}$ at longer times. The corresponding experimental results are shown in Figure \ref{non_rot_1}(a–d), illustrating the features described above. These collective behaviours strongly pointed to anisotropy in the system: the morphology of the trapped cluster, together with the dynamical signatures, indicated coupling between the transverse ($x$) and axial ($z$) degrees of freedom. This is further supported by the reconstructed potential landscape, which explicitly revealed 
$xz$-coupling, while the axial trajectories exhibited intermittent, abrupt displacements (“kicks”) with magnitudes far exceeding typical Brownian fluctuations. To rationalise these observations along two directions, a phenomenological model using two dimensional Langevin equation was developed. The particle was approximated as an ellipsoid, naturally introducing mobility coupling between transverse and axial degrees of freedom, and was confined in a coupled harmonic potential (the exact mathematical expression given in Figure \ref{non_rot_1}). Crucially, an additional force term ($f_{\text{add}}(t)$) was incorporated to represent the observed abrupt kicks: their magnitudes are sampled from a Gaussian distribution with a nonzero mean (introducing drift) and variance proportional to an extra temperature $\delta T$. Consequently, the particle in the axial direction experienced an effective temperature of $T+\delta T$. The waiting times between successive kicks were sampled from a normalised exponential distribution, characterised by a particle-specific kick rate. The dynamics of the particle were governed by the following set of equations:
\begin{equation}
    \begin{aligned}
        m \ddot{x}(t) &= -(\gamma_{xx} \dot{x}(t) + \gamma_{xz} \dot{z}(t)) - k_x x(t) + \alpha z(t) + \xi_{v_x}(t)\\
        m \ddot{z}(t) &= -(\gamma_{zx} \dot{x}(t) + \gamma_{zz} \dot{z}(t)) - k_z z(t) + \alpha x(t) + \xi_{v_z}(t) +\\ & \hspace{1cm} f_{\text{add}}(t),
    \end{aligned}
    \label{xz_eq}
\end{equation}
This phenomenological model successfully reproduced the experimental features of the ACF$_x$, MSD$_x$  and MSD$_z$
, as well as the bimodal probability distribution observed along the axial direction—a prominent feature also observed experimentally (as shown in Figure \ref{non_rot_1}(i-l)). The emergence of bimodality strongly suggested active-like behaviour in the $z$-direction. Moreover, simulations revealed that the kick rate scaled directly with laser power, providing a direct means to tune the level of activity along the axial degree of freedom (Figure \ref{non_rot_1}(g-h)).

\section*{Applications}

As fundamental as this force is, being responsible for processes such as planet formation\cite{teiser2013photophoresis,herrmann2007effects,cuello2016effects, wurm2013photophoretic, mousis2007photophoresis, smallwood2015photophoresis, moudens2011photophoretic} and enabling numerous fundamental studies through particle trapping, as discussed in the preceding sections, photophoretic trapping has also found wide-ranging practical applications, which will be discussed in the following sections. 

\subsection*{3D Volumetric display}

Three-dimensional (3D) volumetric displays aim to create images that occupy real space and can be viewed from nearly any direction without the need for stereoscopic glasses or head tracking. Unlike planar displays, which render 3D content through binocular disparity or computational light-field synthesis, volumetric displays generate light from actual points distributed throughout a physical volume. Existing optical approaches include holography~\cite{ruiz2016holovect}, swept-volume displays~\cite{gately2011three}, and light-field projection~\cite{hirsch2014compressive} but these methods often suffer from viewing-angle restrictions, image clipping, limited refresh rates, or the inability to form true free-space images.

A major advance was achieved with the introduction of a photophoretic-trap-based volumetric display~\cite{smalley2018photophoretic, barton2021photophoretic}, in which a single micron-sized cellulose particle was trapped using a 405~nm beam shaped by spherical and astigmatic aberrations to create a stable photophoretic trap. By rapidly scanning the particle through a three-dimensional region and illuminating it sequentially with red, green, and blue lasers, sub-10~\(\mu\)m voxels were generated with nearly \(360^\circ\) viewing angles and colour performance surpassing holographic and light-field systems.

Photophoretic volumetric displays (as shown in top figure of figure \ref{future_directions} in left panel) provide several key advantages: they produce real physical voxels suspended in air, eliminate occlusion and viewing-direction constraints, and yield speckle-free, wide-gamut colour images without computational reconstruction. Recent work using metallic particles in a ``boat trap’’ configuration has improved trapping stability and predictability~\cite{mirzaei2024optical}, while trapping with multimode fibres offers higher photophoretic forces suitable for larger display volumes in free-space~\cite{sil2024ultrastable}. 

Looking forward, progress is expected through parallel trapping of multiple particles using spatial light modulators~\cite{porfirev2016dynamic}, faster beam steering via acousto-optic or electro-optic deflectors, and the development of engineered particles with improved stability, brightness, and colour response. Such advances could enable practical, real-time, full-colour volumetric displays for visualization and communication as has been already envisaged in popular science fiction and also interactive medical interfaces. 

\subsection*{Spectroscopy}

The development of photophoretic spectroscopy originated with the pioneering work of Pope, Arnold, and Rozenshtein, who demonstrated that the wavelength dependence of the photophoretic force acting on a single levitated micron-sized perylene crystallite reproduced its optical absorption spectrum, establishing photophoresis as a viable probe of absorption in individual particles~\cite{pope1979photophoretic}. This was followed by Arnold, Amani, and Orenstein, who constructed the first dedicated \emph{photophoretic spectrometer}, enabling continuous force-based absorption measurements from 200--1000\,nm at pressures down to $10^{-4}$\,Torr and revealing clear correspondence between photophoretic-force spectra and the CdS band edge absorption~\cite{arnold1980photophoretic}. Soon after, Arnold and Amani extended the method to broadband spectroscopy of individual CdS crystallites, using reduced pressures and electronic damping to improve sensitivity and suppress radiometric-torque–induced oscillations~\cite{arnold1980broadband}. A related technique was introduced by Lin, who used radiometric and photophoretic forces in a quadrupole trap to obtain mid-infrared absorption spectra (930--1080\,cm$^{-1}$) from single ammonium sulfate particles, marking the first vibrational photophoretic spectra and showing excellent agreement with FTIR measurements~\cite{lin1985infrared}. 

With the advent of stable free-space photophoretic traps, photophoretic spectroscopy evolved from a force-measurement tool into an excitation pathway for \emph{Raman spectroscopy}. Pan \textit{et al.}~\cite{pan2012photophoretic} demonstrated the first Raman spectra (as shown in middle figure of figure \ref{future_directions} in left panel)  of absorbing particles stably trapped in air using counter propagating hollow beams, enabling analysis of carbonaceous particles without substrates or liquid media. This method was significantly expanded by Wang \textit{et al.}, who established photophoretic trapping–Raman spectroscopy (PTRS) for both strongly absorbing and weakly absorbing bioaerosols including pollens and fungal spores, achieving single-particle Raman spectra in the 1600–3400\,cm$^{-1}$ region and highlighting its ability to interrogate chemically complex biological materials directly in air~\cite{wang2015photophoretic}.

Compared to traditional absorption or Raman  spectroscopy,  photophoretic spectroscopy offers several key advantages: (i) true single-particle sensitivity without ensemble averaging; (ii) compatibility with opaque, irregular, or strongly scattering particles; (iii) substrate-free and contamination-free measurements in air and (iv) simultaneous access to particle mass, charge, morphology, and optical absorption. The PTRS variants further eliminate the need for optical transparency required in radiation-pressure (optical tweezer) Raman spectroscopy\cite{dai2021optical}, enabling studies of particles that conventional optical traps cannot hold. Nonetheless, limitations remain: the requirement for stable levitation (often at reduced pressure), sensitivity to particle orientation, the need for sufficiently absorbing particles to generate usable photophoretic forces, and limited signal strength for weak Raman scatterers. Despite these constraints, photophoretic spectroscopy and PTRS represent a uniquely powerful set of techniques for characterizing the optical, chemical, and physical properties of individual airborne particles, spanning semiconductors, carbonaceous aerosols, and complex bioaerosols. 

\subsection*{Macroscale Photophoretic Flight and Near-Space Applications}

\begin{figure*}[h]
    \centering
    \includegraphics[width=0.7\textwidth]{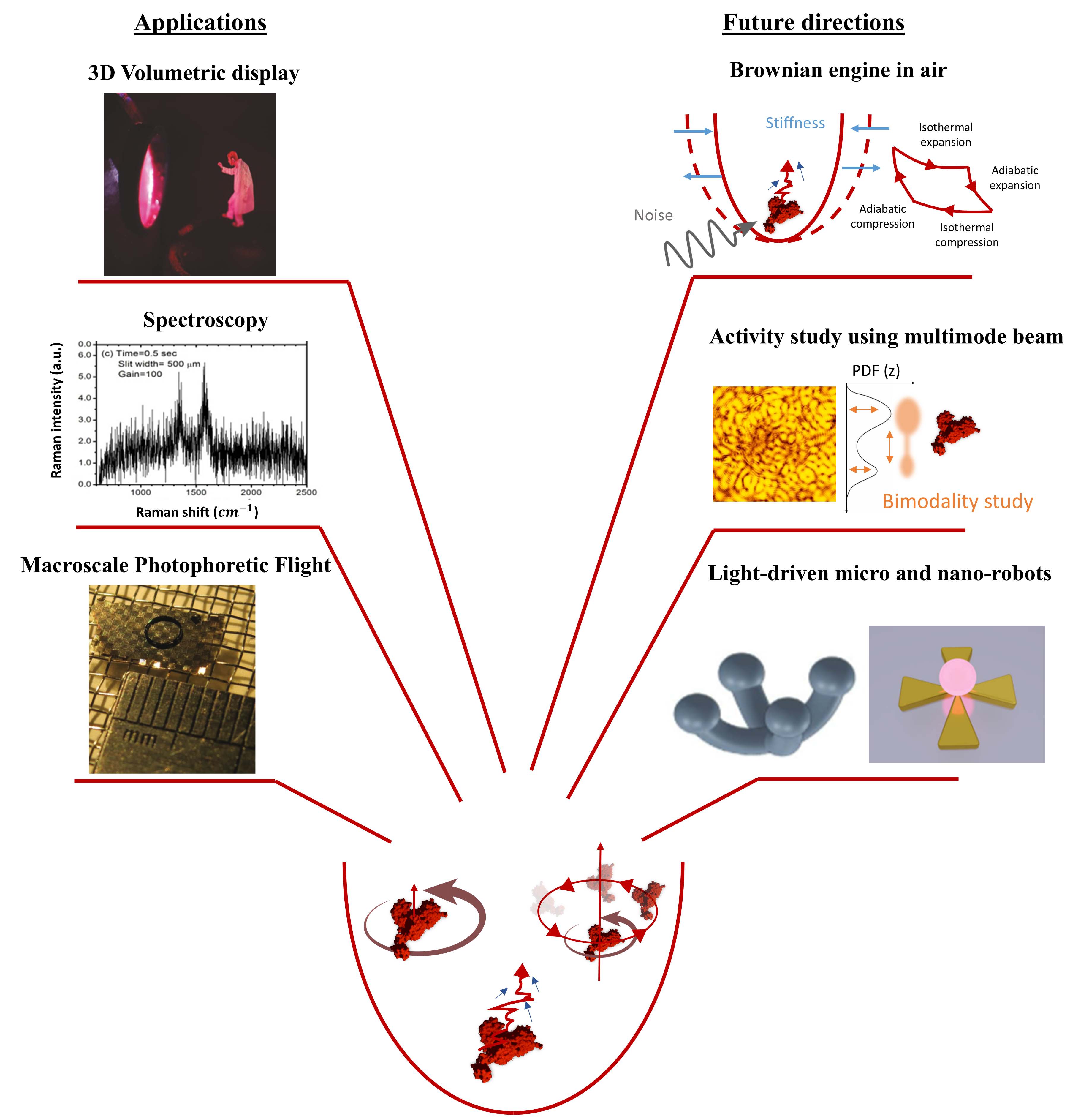}
    \caption{Applications and future directions of photophoretic trapping- \textit{Left panel}: Demonstrated applications of photophoretic trapping, including macroscale photophoretic flight~\cite{cortes2020photophoretic} \copyright \ 2020 Wiley-VCH GmbH, Raman Spectroscopy of absorbing particles ~\cite{pan2012photophoretic} \copyright \ 2012 Optica Publishing Group, and 3D volumetric display based on photophoretic manipulation ~\cite{smalley2018photophoretic} \copyright \ 2018 Springer Nature. \textit{Right panel}: Potential future research directions include realizing high-efficiency Brownian engines operating in air, investigating particle activity under multimode optical fields \cite{sil2024ultrastable} \copyright \ 2024 American Chemical Society, and developing controlled micro- and nanorobots  \cite{qin2025light, zhang2020distributed} \copyright \ 2025 Springer Nature \copyright \ 2020 Wiley-VCH GmbH.}
    \label{future_directions}
\end{figure*}

The extension of photophoretic forces from microscopic aerosols to macroscopic, centimeter-scale structures has led to a new class of light-driven flyers capable of sustained levitation in rarefied environments. Initial advances showed that ultralight, internally structured plates constructed from hollow ceramic membranes containing dense arrays of microchannels can generate strong thermal transpiration lift under modest illumination (as shown in bottom figure of figure \ref{future_directions} in left panel) . These microchannel architectures enable creep-flow propulsion that significantly exceeds conventional $\Delta T$ and $\Delta \alpha$ photophoretic mechanisms, allowing millimeter to centimeter scale objects to hover stably at atmospheric pressure and to freely levitate at reduced pressure~\cite{cortes2020photophoretic}. Subsequent developments demonstrated that lithography-patterned and additively manufactured ultrathin flyers with engineered thermal asymmetry can also achieve controlled hovering, rotation, and free-flight under very low optical intensities. These studies established that macroscopic photophoretic lift can be effectively tuned through structural geometry, thermal emissivity, channel arrangement, and gas surface interactions~\cite{celenza2024three}.  

More recent work has provided a comprehensive analytical and experimental framework describing how perforated dual-membrane structures can harness thermal transpiration to support centimeter-scale levitation under illumination intensities comparable to sunlight. Optimized membranes combining tailored porosity, patterned ligaments, and selective optical coatings were shown to generate sufficient photophoretic lift to overcome gravity at mesospheric pressures, achieving levitation at intensities below $1~\mathrm{kW\,m^{-2}}$ and supporting payloads ranging from tens to hundreds of milligrams~\cite{schafer2025photophoretic}. These results further enabled concept designs for photophoretic aircraft with characteristic sizes of $3$--$80$\,cm capable of persistent daytime flight at altitudes of $60$--$80$ km, offering opportunities for low-cost atmospheric sensing, passive communication relays, and autonomous exploration of rarefied environments. While challenges remain such as ensuring structural robustness, mitigating nighttime descent, and developing materials with high thermal contrast and resilience, the emerging body of research firmly establishes thermal-transpiration photophoresis as a scalable and energy-efficient approach to sunlight-powered flight in the upper atmosphere. 

\section*{Future Directions}

\subsection*{Complete theoretical understanding of photophoretic force-based optical trapping}

One of the biggest lacunae in research on photophoretic force-based optical trapping is the lack of a first-principles based complete theoretical model that explains such trapping as an emergent phenomena of the various diverse physical processes involved. In our sections on the theoretical premise, most of the formulae provided are rather emperical, and do not have analytical treatment. Indeed, even the reason why photophoretic traps are able to confine particles in the direction transverse to that of the light propagation is not understood quantitatively as of now. In addition, the exact role of asymmetry -- both in terms of surface morphology and thermal accommodation is not clearly spelled out. Most importantly, while it is expected that the interaction of the radial component of the photophoretic force and gravity would lead to rotation of a particle as a necessary consequence of being trapped, observations by Pahi et al. \cite{pahi2024study,pahi2025light} have clearly enunciated that not all particles need to rotate to facilitate stable trapping. The authors do not provide a theoretical model to explain their observations, and there exists a substantial research gap in this area. The random axial jumps observed in asymmetric particles \cite{sil2024ultrastable,pahi2025light} has also not been quantitatively explained. Indeed these stochastic jumps may well prove an issue in the development of efficacious volumetric displays with photophoretically trapped particles, so that their understanding and possible mitigation is of considerable importance.

\subsection*{Facilitator of Non-Equilibrium Statistical Mechanics phenomena at the mesoscopic scale in air}

The fact that a photophoretic trapping system operates in air, coupled with the nature of the trapping process naturally places it in a non-equilibrium regime: thus, a trapped particle is continuously driven by an external energy flux, exhibits complex dynamical motion while confined, and retains inertial effects owing to the low viscosity of the medium. This setting provides a unique opportunity to investigate statistical mechanics and thermodynamic behavior, particularly in the context of entropy production, thereby enriching the broader understanding of non-equilibrium processes in Brownian systems in varied environments. Indeed, the non-equilibrium dynamics observed by Pahi et al. \cite{pahi2025light}, particularly the coexistence of active and diffusive motion in photophoretically trapped microclusters\cite{pahi2025light}, provides a natural platform for realizing Brownian engines in air. Since the longitudinal dynamics shows active behaviour which can be tuned through the trapping beam power, the system offers direct experimental control over the degree of activity. This tunability, combined with the inherently non-equilibrium nature of the dynamics, makes it possible to construct Brownian engines that operate at efficiencies potentially exceeding the Carnot bound, as predicted in active matter systems \cite{holubec2020active}. Future work will involve designing cyclic protocols to harness these dynamics for work extraction, and investigating the thermodynamic aspects of trajectory-level energetics (work, heat, entropy production)\cite{das2022inferring}, thereby paving the way toward high-efficiency active microscopic engines operating in gaseous environments (a schematic shown in top figure of figure \ref{future_directions} in right panel).  In addition, while current works have primarily employed a Gaussian trapping beam to investigate particle dynamics, the use of engineered complex beam profiles, such as that from multimode fibers\cite{sil2024ultrastable} offer an exciting avenue to enhance the active behaviour of trapped particles (a schematic shown in middle figure of figure \ref{future_directions} in right panel). 

\subsection*{Light-driven micro and nano-robots}
Previous works has used  optical tweezers to create and manipulate micro/nano robots \cite{zhang2020distributed, qin2025light}. Photophoresis, as shown in the seminal works of Shvedov et al. \cite{shvedov2009optical} can generate forces leading to giant manipulation of microparticles. Presently, the knowledge that particle asymmetry may be used as a handle to drive exotic dynamics \cite{pahi2024study} -- both rotational and translational -- can add further to this capability and lead to the engineering of designed morphologies that can prove to be efficacious for propulsion or steering mechanisms for untethered micro and nano-robots operating in low-pressure environments or sealed microfluidic chambers, as well as in aerosol navigation. 

\subsection*{Low-Pressure Space and Planetary Applications}
Photophoretic-force based trapping warrants investigation in space-relevant pressures, where photophoretic effects are considerable, enabling exciting applications or even new methods for dust mitigation, sample handling, and autonomous particle transport in extraterrestrial environments. Indeed, future research may also address light-based propulsion for micro/nano-satellites, CubeSats and the growth of planetesimals.

\subsection*{Aerosol Manipulation, Separation, and Spectroscopy}
There exist very interesting opportunities to design next-generation aerosol filters and particle sorters that exploit thermal transpiration to selectively trap, concentrate, or classify nanoparticles based on their optical and thermal properties. More specifically, the advent of single\cite{sil2022trapping} and multimode-fiber\cite{sil2024ultrastable} based photophoretic trapping mechanisms can lead to the development of hand-held devices which may be used in-situ for both fluorescence and Raman spectroscopy - something which may create new paradigms for aerosol detection, and even manipulation.

\section*{Conclusions}

Photophoretic trapping has matured into a powerful strategy for manipulation in gaseous media, unifying optical, thermal, and fluid-dynamic principles in a way that no other trapping mechanism achieves. Across the studies surveyed in this review, a clear trajectory emerges: from foundational descriptions of photophoretic forces to trapping geometries and rich dynamics of trapped particles, and an expanding suite of applications in the direction of spectroscopy, aerosol science, display technologies, and even macroscale flight in rarefied atmospheres. This underscores the flexibility of photophoretic systems and their unique suitability for environments where conventional optical forces are weak or ineffective. Continued advances in particle design, beam shaping, and thermal engineering promise to further enhance control, stability, and scalability of photophoretic manipulation. As these developments converge, photophoretic trapping stands to play an increasingly important role in both fundamental research and the realization of novel technologies that leverage light-driven motion in free space.

\section*{Acknowledgements}

The authors acknowledge Indian Institute of Science Education and Research (IISER) Kolkata for the funding and facilities.

\section*{Conflict of Interest}

The authors declare no conflicts of interest

\begin{shaded}
\noindent\textsf{\textbf{Keywords:} \keywords} 
\end{shaded}


\setlength{\bibsep}{0.0cm}

\clearpage


\end{document}